%% file: root.tex
\PassOptionsToPackage{compatibility=false}{caption}

\documentclass[letterpaper, 10 pt, conference]{ieeeconf}
\IEEEoverridecommandlockouts
\usepackage[utf8]{inputenc}
\usepackage[american]{babel}
\usepackage{csquotes}

\usepackage{comment}
\usepackage{url}
\usepackage{graphicx}
\usepackage{subcaption}
\usepackage{multicol}
\usepackage{xspace}
\usepackage{amsmath, amssymb, amsfonts}
\usepackage{ntheorem}
\usepackage{mathtools}
\usepackage{multirow}
\usepackage{tabularx}

\newcolumntype{C}{>{\centering\arraybackslash}X}
\newcolumntype{L}{>{\raggedright\arraybackslash}X}
\usepackage{siunitx}
\usepackage{listings}
\usepackage{float}

\usepackage{pgfplots}
\usepgfplotslibrary{fillbetween}
\pgfplotsset{compat = newest}
\usetikzlibrary{arrows,positioning,shapes,intersections,patterns,calc,fit,external,decorations,decorations.markings} 
\newtheorem{definition}{Definition}
\newtheorem{lemma}{Lemma}
\newtheorem{theorem}{Theorem}

\newtheorem{remarkth}[definition]{Remark}
\newenvironment{remark}{\begin{remarkth}\upshape}{\end{remarkth}}

\newcommand{\N}{\mathbb{N}}

\usepackage{tikz}
\usetikzlibrary{matrix,arrows,calc,positioning,shapes,decorations.pathreplacing}
\tikzstyle{vertex}=[circle,fill=black!20,minimum size=15pt,inner sep=0pt]
\tikzstyle{selected vertex} = [vertex, fill=red!24]
\tikzstyle{edge} = [draw,thick,-]
\tikzstyle{dedge} = [draw,thick,<->]
\tikzstyle{shadowdedge} = [draw, dotted,->]
\tikzstyle{weight} = [font=\small]
\tikzstyle{selected edge} = [draw,line width=5pt,-,red!50]
\tikzstyle{ignored edge} = [draw,line width=5pt,-,black!20]
\pgfplotsset{select coords between index/.style 2 args={
    x filter/.code={
        \ifnum\coordindex<#1\fi
        \ifnum\coordindex>#2\fi
    }
}}
\usepackage{booktabs}
\input{mydefs.tex}

\begin{document}

\title{\LARGE \bf Learning Rigidity-based Flocking Control with Gaussian Processes}

\author{Manuela Gamonal$^{1}$, Thomas Beckers$^{2}$, George J. Pappas$^{2}$, Leonardo J. Colombo$^{3}$,%
\thanks{$^{1}$ M. Gamonal is with the Instituto de Ciencias Matemáticas (CSIC-UAM-UCM-UC3M), Spain. email:  \{{\tt\small manuela.gamonal@icmat.es},}%
    \thanks{$^{2}$ T. Beckers and G. Pappas are with Department of Electrical and Systems Engineering, University of Pennsylvania, Philadelphia, PA 19104, USA \{{\tt\small tbeckers, pappasg\}@seas.upenn.edu},}%
    \thanks{$^{3}$ L. Colombo is with the Centre for Automation and Robotics
(CSIC-UPM), Ctra. M300 Campo Real, Km 0,200, Arganda
del Rey - 28500 Madrid, Spain, {\tt\small leonardo.colombo@car.upm-csic.es}}
}
\maketitle
\thispagestyle{empty}
\pagestyle{empty}

\maketitle
 \thispagestyle{empty}
 \pagestyle{empty}

\begin{abstract} Flocking control of multi-agents system is challenging for agents with partially unknown dynamics. This paper proposes an online learning-based controller to stabilize flocking motion of double-integrator agents with additional unknown nonlinear dynamics by using Gaussian processes (GPs). Agents interaction is described by a time-invariant infinitesimally minimally rigid undirected graph. We provide a decentralized control law that exponentially stabilizes the motion of the agents and captures Reynolds boids motion for swarms by using GPs as an online learning-based oracle for the prediction of the unknown dynamics. In particular the presented approach guarantees a probabilistic bounded tracking error with high probability. 
\end{abstract}


\section{Introduction}\label{sec1}

Flocking, swarming, and schooling are common emergent
collective motion behaviors exhibited in nature. These natural collective behaviors can be leveraged in multirobot systems to safely transport large cohesive groups of robots within a workspace~\cite{rubenstein2014}. To capture these effects, Reynolds introduced three heuristic rules: cohesion; alignment; and
separation, to reproduce flocking motions in computer graphics in~\cite{reynolds1987flocks}. Decentralized flocking control algorithms based on Reynolds's rules have gained much attention in the recent years due to the increasing amount of mobile and aerial robotic swarms~\cite{tanner2003stable, anderson2, sun2015rigid}.

Decentralized flocking stabilization techniques can be based on artificial potential fields \cite{tanner2003stable}, negative gradient of potential functions~\cite{dika} and rigidity theory~\cite{anderson2, sun2015rigid}. For the latter one, most of techniques to stabilize a set of agents to a desired rigid shape are given for single integrator agents~\cite{krick2009stabilisation}, \cite{oh2015survey}. Double integrator models have been extensively studied for flocking control since ~\cite{olfati2006flocking} and~\cite{tanner2003stable}. In particular rigidity-based flocking control has been studied in \cite{anderson2} and \cite{sun2015rigid}, but non of them consider uncertainties in the model. 

The design of safe controllers for robotic swarms is a substantial aspect for an increasing range of application domains. However, parts of the robot's dynamics and external disturbances are often unknown or very time-consuming to model. 
To overcome the issue of unknown dynamics, learning-based control laws have been proposed but they are limited to iterative learning~\cite{liu2012iterative}, leader-follower formations~\cite{yuan2017formation,yan2021flocking,8870252} or imitation~\cite{schilling2019learning}.  We present a safe decentralized controller for agents with second-order dynamics by using Gaussian Processes, for an online learning of the unknown dynamics, while agents exhibit Reynolds rules of flocking. The presented learning-based controller guarantees a probabilistic bounded error to desired flocking motions with high probability and where the bound can be explicitly given by solving an optimization problem over the data-set. The online learning approach allows to improve the model and, thus, the stability performance during run-time.

Decentralized machine learning control algorithms for multi-agent systems has been recently studied in~\cite{tolstaya2020learning, gama2021graph} using graph neural networks and also in \cite{yang2021distributed, Beckers2021}. In particular recent learning-based methods for flocking control of second-order agents can be found in \cite{Ji2021,schilling2019learning, be2021}, but none of them includes 3D flocking control with stability guarantee in the performance. To the best of the authors' knowledge, there are no available results for the design of a learning-based flocking control law for double integrator agents under partially unknown dynamics, based on online learning data-driven models with exponential stability guarantees.
  
In data-driven control, data of the unknown system dynamics is collected and used to predict the dynamics in areas without training data. In contrast to parametric models, those models are highly flexible and are able to reproduce a large class of different dynamics~\cite{hou2013model}. Recently, Gaussian process (GP) models~\cite{rasmussen2006gaussian} has been increasingly used for modeling dynamical system due to many beneficial properties such as the bias-variance trade-off and the strong connection to Bayesian statistics. In contrast to many other techniques, GP models provide not only a prediction but also a measure for the uncertainty of the model. This powerful property makes them very attractive for many applications in control, e.g., model predictive control~\cite{hewing2019cautious}, feedback linearization~\cite{umlauft:TAC2020}, and tracking control~\cite{beckers2019stable}, as the uncertainty measure allows to provide performance and safety guarantees. The purpose of this article is to employ the power of learning-based approaches, in particular, GPs, for decentralized flocking control of second-order agents with partially unknown dynamics guaranteeing the probabilistic boundedness of the error to desired flocking motions, with high probability.

The remaining article is as follows: we introduce rigid and connected graphs on \cref{sec2}. After the problem setting in~\cref{sec3}, the online learning-based controller is given in~\cref{sec4}. Numerical examples are shown in~\cref{sec5}.

\textbf{Notation:} Matrices are denoted with capital letters. ~$\mathcal{N}(\mu,\Sigma)$ describes a normal distribution with mean~$\mu$ and covariance~$\Sigma$. The probability function is denoted by $\Prob$. $\mathbb{E}[X]$ denotes the expected value of a random variable $X$. $\R_{>0}$ denotes the set of positive real numbers. The Euclidean norm is denoted by $||\cdot||$ and by $|\mathcal{X}|$ the cardinal of the set $\mathcal{X}$.


\section{Agents in a network}\label{sec2}
Consider $n\geq 2$ autonomous agents whose positions are denoted by $\bm{q}_i\in\R^{d}$, $d=\{2,3\}$ and denote by $\bm{q}\in\R^{dn}$ the stacked vector of agents' positions. Neighbor's relationships are described  by an undirected and time-invariant graph $\mathbb{G} = (\mathcal{N}, \mathcal{E})$ with the ordered edge set $\mathcal{E}\subseteq\mathcal{N}\times\mathcal{N}$. The set of neighbors for $i\in\mathcal{N}$, is defined by $\mathcal{N}_i:=\{j\in\mathcal{N}:(i,j)\in\mathcal{E}\}$. Agents can sense the relative positions of its nearest neighbors, in particular, agents can measure its Euclidean distance from other agents in the subset $\mathcal{N}_i \subseteq \mathcal{N}$. We define the elements of the incidence matrix $B\in\R^{|\mathcal{N}|\times|\mathcal{E}|}$ that establish the neighbors' relationships for $\mathbb{G}$ by $\displaystyle{
	b_{i,k} = \begin{cases}+1  &\text{if} \quad i = {\mathcal{E}_k^{\text{tail}}} \\
		-1  &\text{if} \quad i = {\mathcal{E}_k^{\text{head}}} \\
		0  &\text{otherwise}
	\end{cases}}$, where $\mathcal{E}_k^{\text{head}}$ and $\mathcal{E}_k^{\text{tail}}$  denote the head and tail nodes, respectively, of the edge $\mathcal{E}_k$, i.e., $\mathcal{E}_k = (\mathcal{E}_k^{\text{head}}, \mathcal{E}_k^{\text{tail}})$. The stacked vector of relative positions between neighboring agents, denoted by $\bm{z}\in\R^{d|\mathcal{N}|}$, is given by $\bm{z} = \overline B^T\bm{q}$, where $\overline B := B \otimes I_{d}\in\R^{d|\mathcal{N}|\times d|\mathcal{E}|}$ with $I_{d}$ being the $(d\times d)$ identity matrix, and $\otimes$ the Kronecker product. Note that $\bm{q}_k \in \R^d$ in $\bm{q}$ corresponds to $\bm{q}_i - \bm{q}_j$ for the edge $\mathcal{E}_k$. We define $\bm{z}_{k} := \bm{q}_i - \bm{q}_j$ to simplify notation. A \textit{framework} for $\mathbb{G}$ is defined as the pair $(\mathbb{G},\bm{q})$. 

Flocking control can be achieved by means of the Laplacian matrix associated with $\mathbb{G}$. The Laplacian matrix $L$ is the matrix whose entries are given by $l_{ij}=-1$ for $i\neq j$, if there is an edge between agents $j$ and $i$, else $l_{ij}=0$. Moreover, $\displaystyle{l_{ii}=-\sum_{j\in\mathcal{N}_{i}}l_{ij}}$.  In the case of $\mathbb{G}$ being an undirected graph, it follows that $L=BB^{T}$ (see \cite{godsil2001algebraic} for instance). For a connected and undirected graph there holds $\hbox{rank}(L)=|\mathcal{N}|-1$ and $\hbox{null}(L)=\hbox{null}(B)=\{\mathbf{1}_{|\mathcal{N}|}\}$. Therefore, $\hbox{null}(L)=\hbox{null}(B)=\{\mathbf{1}_{|\mathcal{N}|}\otimes I_{d\times d}\}$. Thus $L$ is positive definite by restricting it to the vector space of $\hbox{span}(\mathbf{1}_{|\mathcal{N}|}\otimes I_{d\times d})^{\perp}$ and the number of connected components in the graph equals the algebraic multiplicity of its null eigenvalue. Besides, the second smallest eigenvalue of $L$, denoted by $\lambda_{2}$, is known as algebraic connectivity of $\mathbb{G}$, as it is related with the interconnection of the nodes.

To guarantee convergence to desired shapes we consider the \textit{rigidity} of the desired formation shapes. The \textit{rigidity matrix} for the framework $(\mathbb{G},\bm{q})$ is defined as, see \cite{asimov}, 
\begin{align}
    R(\bm{z})=\frac{1}{2} \frac{\partial \ell_{\mathbb{G}}(\bm{q})}  {\partial \bm{q}} = D(\bm{z})^{\top} \overline{B} \in \R^{|\mathcal{E}| \times d|\mathcal{N}|},
\end{align}
 with $D(\bm{z}) = \diag(\bm{z}_{1}, \ldots,\bm{z}_{|\mathcal{E}|}) \in \R^{d|\mathcal{E}| \times |\mathcal{E}|}$ and distance measure function $\ell_{\mathbb{G}}: \R^{d|\mathcal{N}|} \rightarrow \R^{|\mathcal{E}|}$ defined by $\displaystyle{
     \ell_{\mathbb{G}}(\bm{q})=\left(\left\|\bm{q}_{i}-\bm{q}_{j}\right\|^{2}\right)_{(i, j) \in \mathcal{E}}=D^\top(\bm{z})\bm{z}}$. Denote the desired distance between neighboring agents over the edge $\mathcal{E}_k$ as $d_{k}$ and define the squared distance error for $\mathcal{E}_k$ as
\begin{align}\label{e_k}
    \bm{e}_{k}=\left\|\bm{q}_{i}-\bm{q}_{j}\right\|^{2}-d_{k}^{2}=\left\|\bm{z}_{k}\right\|^{2}-d_{k}^{2},
\end{align} 
with stacked squared distance vector error denoted by $\bm{e} = \left[\bm{e}_{1}, \ldots, \bm{e}_{|\mathcal{E}|}\right]^{\top}$.%

A framework $(\mathbb{G},\bm{q})$ is said to be \textit{rigid} if it is not possible to smoothly move one node of the framework without moving the rest while maintaining the inter-agent distance given by $\ell_{\mathbb{G}}(\bm{q})$, see~\cite{asimov}. An \textit{infinitesimally rigid} framework is a rigid framework which is invariant under and only under infinitesimally transformations under $R(\bm{z})$, i.e., $\ell_{\mathbb{G}}(\bm{q}+\delta \bm{q})=\ell_{\mathbb{G}}(\bm{q})$ where $\delta \bm{q}$ denotes an infinitesimal displacement of $\bm{q}$.  It is well known (see \cite{asimov} for instance) that a framework  $(\mathbb{G},\bm{q})$ is infinitesimally rigid in $\R^d$ if $\bm{q}$ is a regular value of  $\ell_{\mathbb{G}}(\bm{q})$ and $(\mathbb{G},\bm{q})$ is rigid in $\R^d$. In particular, $(\mathbb{G},\bm{q})$ is \textit{infinitesimally rigid} in $\R^2$ if $\rank R(\bm{z})=2n-3$ (respectively,  $\rank R(\bm{z})=3n-6$ in $\R^3$).

The framework $(\mathbb{G},\bm{q})$ is said to be \textit{minimally rigid} if it has exactly $2n-3$ edges in $\R^2$ or $3n-6$ edges in $\R^{3}$. This means that if we remove one edge from a minimally rigid framework $(\mathbb{G},\bm{q})$, then it is not rigid anymore. Thus, the only motions over the agents in a minimally rigid framework, while they are already in the desired shape, are the ones defining translations and rotations of the whole shape, see \cite{oh2015survey}. One important property for the stabilization to desired motions in shape control with flocking behaviour \cite{anderson2} is that the rigidity matrix $R(\bm{z})$ has full row rank if $(\mathbb{G},\bm{q})$ is minimally and infinitesimally rigid. 

\section{Modeling flocking control of double integrator agents with Gaussian Processes}\label{sec3}
Consider the set $\mathcal{N}$ consisting of $n\geq 2$ free autonomous agents evolving on $\R^d$, $d=\{2,3\}$, as in Section \ref{sec2}, under a double integrator dynamics, that is  $\begin{cases}
	\dot{\bm{q}}&=\bm{v} \\
	\dot{\bm{v}}&=\bm{u}.
	\end{cases}$

According to the Reynolds flocking model \cite{reynolds1987flocks}, the motion of every agent in the flock is defined by the three rules of \textit{alignment, cohesion and separation}. Cohesion and separation might be achieved by  using artificial potential fields \cite{tanner2003stable}. We are interested on shape control with flocking motion, that is, agents reach a desired formation shape and they also move along the $2$D plane or $3$D space by achieving a consensus on their velocities. To exponentially achieve this collective behaviour, and provide the their convergence rates, similarly as in \cite{anderson2} and \cite{sun2015rigid}, we  make the following assumption 

\begin{assum}\label{ass:0}

$(\mathbb{G},\bm{q})$ is an infinitesimally and minimally rigid framework with $\mathbb{G}$ undirected, static and connected.
\end{assum}

To reach a desired shape, for each edge $\mathcal{E}_k=(i,j)$ in the infinitesimally and minimally rigid framework we introduce the artificial potential functions $V_{k}:\R^{d}\to\R$, given by $V_{k}(\bm{z}_k)=\frac{1}{4}(||\bm{z}_{k}||^2-d_{k}^2)^2$, to provide a measure for the interaction between agents and their nearest neighbors (see \cite{oh2015survey} for a detailed discussion on the choices of elastic potential functions).  In these potentials,~$\bm{z}_k$ denotes the relative position between agents for the edge $\mathcal{E}_k$, and $d_{k}$ denotes the desired length for the edge $\mathcal{E}_k$. 

The acceleration of agent $i$ is determined by 
$$\bm{u}_i(t)=-l_{ij}\underbrace{\sum_{j\in\mathcal{N}_{i}}(\bm{v}_i-\bm{v}_j)}_{\hbox{alignment}}-\underbrace{\sum_{j\in\mathcal{N}_{i}}\nabla_{\bm{q}_i}V_{k},}_{\hbox{cohesion + separation}}$$ that is, for $\mathcal{L}=L\otimes I_d$, then $ \bm{u}(t)=-\mathcal{L}\bm{v}-R^{T}(\bm{z})e(\bm{z})$.

The closed loop system, called \textit{double integrator flocking stabilization system} \cite{anderson2,sun2015rigid} is given by 
\begin{align}\label{formation}
	\begin{cases}
	\dot{\bm{q}}_i =\bm{v}_i \\
	\dot{\bm{v}}_i = -l_{ij}\displaystyle{\sum_{j\in\mathcal{N}_{i}}(\bm{v}_i-\bm{v}_j)-\sum_{j\in\mathcal{N}_{i}}\nabla_{\bm{q}_i}V_{k}}.
	\end{cases}
\end{align}

Therefore, we can define the artificial potential function $V_0:\R^{d|\mathcal{N}|}\to\R$ for the overall networked control system as 
\begin{align}
    V_0(\bm{z})=\sum_{k=1}^{|\mathcal{E}|}V_k(||\bm{z}_k||).
\end{align}
In order to control the velocity of the agents, we introduce the disagreement vector $\bm{\delta}\in\mathbb{R}^{d|
\mathcal{N}|}$. Consider the potential function $V_1:\R^{d|\mathcal{N}|}\to\R$ defined as
\begin{align}
    V_1(\bm{\delta})=\frac{1}{2}\sum_{i=1}^{|\mathcal{N}|}||\bm{\delta}_k||^2,
\end{align} $\bm{\delta}_k=\bm{\delta}_i-\bm{\delta}_j$,  where $\bm{\delta}=[\bm{\delta}_1^{T},\ldots,\bm{\delta}_{|\mathcal{N}|}^{T}]^{T}$ is the velocity disagreement vector where each component $\bm{\delta}_i$ is given by $\bm{\delta}_i=\bm{v}_i-\bar{\bm{v}}$ with $\displaystyle{\bar{\bm{v}}(t):=\frac{1}{|\mathcal{N}|}\sum_{i=1}^{|\mathcal{N}|}\bm{v}_i(t)}\in\mathbb{R}^{d}$ denoting the average velocity of the agents. Note that $\dot{\bar{\bm{v}}}(t)=0$ and hence $\dot{\bm{\delta}}_i=\dot{\bm{v}}_i$, since $\bar{\bm{v}}$ is constant. 

By considering the semi-definite function $V:=V_0+V_1$ as energy function for the double integrator flocking stabilization system \eqref{formation}, under \cref{ass:0}, one can show local exponential convergence of the agents to desired shapes with flocking motion behaviour \cite{anderson2}, \cite{sun2015rigid}.

Next, consider each agent $i\in\{1,\ldots,|\mathcal{N}|\}$ disturbed by an additive unknown dynamics given by \begin{align}\label{double}
    \begin{cases}
      \dot{\bm{q}}_i=\bm{v}_i, \\
      \dot{\bm{v}}_i= \u_i+\f_i(\bm{q}_i,\bm{v}_i),
    \end{cases}       
\end{align}
where $\f_i:\R^{2d}\to\R^{d}$ is a state-dependent unknown function.
The time-dependency of the states is omitted for simplicity of notation and the time dependency of the unknown input forces $\f_i$ might be also indirect, i.e. $\f_i(\bm{q}_i(t),\bm{v}_i(t))$. 

In the following, we propose an online learning strategy for flocking motion, and an upper probabilistic bound for the error estimation between the learned and the true dynamics (i.e., the mean prediction of the GP).
In preparation for the learning and control step, we introduce the estimate $\hat{\f}_i\colon\R^{2d}\to\R^{d}$ which can include existing prior knowledge about the unknown dynamics $\f_i$, e.g, using classical system identification modeling \cite{aastrom1971system}. However, if no prior knowledge is available, the estimate $\hat{\f}_i$ is set to zero. Then, ~\cref{double} can be written as
 \begin{align}\label{double2}
    \begin{cases}
      \dot{\bm{q}}=\bm{v}, \\
      \dot{\bm{v}}=\u+\bm{\rho}(\bm{q})+\hat{\f}(\bm{q},\bm{v}),
    \end{cases}   
\end{align}
with the stacked vector of estimating functions $\hat{\f}(\bm{q},\bm{v})=[\hat{\f}_1(\bm{q}_1,\bm{v}_1)^\top,\ldots,\hat{\f}_{|\mathcal{N}|}(\bm{q}_{|\mathcal{N}|},\bm{v}_{|\mathcal{N}|})^\top]^\top$ and the unknown dynamics $\bm{\rho}\colon\R^{2d|\mathcal{N}|}\to\R^{d|\mathcal{N}|}$ with elements defined by
\begin{align}\label{estimation-error}
    \bm{\rho}_i(\bm{p}_i)=\f_i(\bm{q}_i,\bm{v}_i)-\hat{\f}_i(\bm{q}_i,\bm{v}_i),
\end{align}
where $\bm{p}_i=[\bm{q}_i^\top,\bm{v}_i^\top]^\top$. In the next step, we employ a GP model for the learning of the unknown dynamics $\bm{\rho}$. 

Gaussian processes are stochastic processes which are completely defined by a mean function $m_{\mathrm{GP}}: \R^p \rightarrow \R$, $p\in\mathbb{N}$ and a kernel function $k: \R^p \times \R^p \rightarrow \R$. Then, for all $f_{\mathrm{GP}}(\bm{x}) \sim \mathcal{G} \mathcal{P}\left(m_{\mathrm{GP}}(\bm{x}), k\left(\bm{x}, \bm{x}^{\prime}\right)\right)$, it is verified that the mean $m_{\mathrm{GP}}(\bm{x}) =\mathbb{E}\left[f_{\mathrm{GP}}(\bm{x})\right]$ and the variance $k\left(\bm{x}, \bm{x}^{\prime}\right) =\mathbb{E}\left[\left(f_{\mathrm{GP}}(\bm{x})-m_{\mathrm{GP}}(\bm{x})\right)\left(f_{\mathrm{GP}}\left(\bm{x}^{\prime}\right)-m_{\mathrm{GP}}\left(\bm{x}^{\prime}\right)\right)\right]$, for any $\bm{x}, \bm{x}^{\prime} \in \R^p,p\in\N$. 
One of the main strengths of GPs is that, when combined with Bayes' Theorem, they can provide statistical inference with function regression.

Consider the output $\bm{y}$ of a function $\bm{f}: \R^{p} \rightarrow \R^{p}$. Without loss of generality, the mean function has been set to zero.  Measurements may be affected by Gaussian noise such that $\bm{y} = \bm{f}(\bm{x}) + \bm{\eta}$, where $\bm{\eta}\sim\mathcal{N}(0,\sigma^2 I_p)$ with the $p$-dimensional identity matrix $I_p$. 
The set of input data, $X=\left[\bm{x}^{\{1\}}, \bm{x}^{\{2\}}, \ldots, \bm{x}^{\{m\}}\right] \in \R^{p \times m}$ and measured output data, $Y=\left[\bm{y}^{\{1\}}, \bm{y}^{\{2\}}, \ldots, \bm{y}^{\{m\}}\right] \in \R^{p \times m}$ constitute the training set $ \mathcal{D=} \left\lbrace X,Y \right \rbrace$. Then, for a test input $\bm{x}^{*}\in\R^p$ the predictions of $\bm{f}(\bm{x}^{*})$ are obtained by conditioning on the data which leads to the posterior distribution 
\begin{align}
\mu\left(f_{i}\!\mid\!\bm{x}^{*}, \D\right)&\!=\! \bm{k}\left(\bm{x}^{*}, X\right)^{\!\top}\!\left(K+I \sigma^{2}\right)^{-1} Y_{:, i},\\
    \var\left(f_{i}\mid x^{*}, \D\right)&= k\left(\bm{x}^{*}, \bm{x}^{*}\right)-\bm{k}\left(\bm{x}^{*}, X\right)^{\top}\notag\\
    &\phantom{=}\left(K+I\sigma^{2}\right)^{-1} \bm{k}\left(\bm{x}^{*}, X\right)\notag
\end{align}
for all $i\in\{1,\ldots,p\}$, where $Y_{:,i}$ denotes the $i$-th column of the outputs matrix~$Y$. The kernel $k$ measures the correlation of two inputs~$(\x,\x^\prime)$. The function~$K\colon \R^{p\times m}\times  \R^{p\times m}\to\R^{m\times m}$ is called the Gram matrix and its elements are given by $K_{j',j}= k(X_{:, j'},X_{:, j})+\delta(j,j')\sigma^2$ for all $j',j\in\{1,\ldots,m\}$ with the delta function $\delta(j,j')=1$ for $j=j'$ and zero, otherwise. The vector-valued function~$\bm{k}\colon \R^p\times  \R^{p\times m}\to\R^m$, with elements~$k_j = k(\bm{x}^*,X_{:, j})$ for all $j\in\{1,\ldots,m\}$, expresses the covariance between~$\x^*$ and the input training data $X$. The choice of the kernel and the determination of the corresponding hyperparameters can be seen as degrees of freedom of the regression procedure. One common and powerful kernel for GP models of physical systems is the squared exponential kernel. An overview about different kernels and their properties is provided in~\cite{rasmussen2006gaussian}. To simplify, identical kernels are selected for each output dimension. Nevertheless, the GP model can be easily adapted to different kernels for each output dimension.

For our purpose, each agent collects $m(t)\in\N$ training points based on its own dynamics~\cref{double} such that data sets
\begin{align}
    \D_{i,m(t)}=\{\bm{p}_i^{\{j\}},\y_i^{\{j\}}\}_{j=1}^{m(t)}\label{for:dataset}
\end{align}
are created. The output data $\y_i\in\R^d$ are given by $\y_i=\dot{\bm{v}}_i-\hat{\f}_i(\bm{q}_i,\bm{v}_i)-\u_i$. The number of training points $m(t)$ of the data sets $\D_{i,m(t)},i\in\{1,\ldots,|\mathcal{N}|\}$ with $m\colon \R_{\geq 0}\to\N$ can change over time $t$, i.e., it allows \textit{online learning}. Let $\D_{m(t)}=\{\D_{i,m(t)},\ldots,\D_{|\mathcal{N}|,m(t)}\}$ be a set that contains all training set. We introduce the following assumption on the data collection.
\begin{assum}\label{ass:1}
    There are only finitely many switches of~$m(t)$ over time and there exists a time $T\in\R_{\geq 0}$ where $\D_{m(T)}=\D_{m(t)},\forall t\geq T,\forall i\in\{1,\ldots,|\mathcal{N}|\}$.
\end{assum}

\begin{remark}~\cref{ass:1} ensures that the switching between the data sets is not infinitely fast which is natural in real world applications.  \end{remark}

To model the error, an assumption has to be made about the kernel function $k$ of the GP model.

\begin{assum}\label{ass:2}
    The continuous kernel $k$ is chosen in such a way the functions $\rho_i$, $i\in\{1,\ldots,d|\mathcal{N}|\}$ have a bounded reproducing kernel Hilbert Space (RKHS) norm on a compact set $\Omega \subset \R^{2d|\mathcal{N}|}, \text { i.e. }\left\|\rho_i\right\|_{k}<\infty \text { for all } i\in\{1, \ldots, d|\mathcal{N}|\}$. 
\end{assum}
\begin{remark}The norm of a function in a RKHS is a smoothness measure relative to a kernel~$k$ that is uniquely connected with this RKHS. In particular, it is a Lipschitz constant with respect to the metric of the used kernel ~\cite{wahba1990spline}.\end{remark}

Under the previous consideration on the model, the model error can be probabilistically bounded as written in the following lemma, which is a direct consequence of Lemma $1$ given in~\cite{beckers2017stable}.

\begin{lemma}\label{model-error}Consider the system \cref{double2} and a GP model satisfying~\cref{ass:1,ass:2}. Then the model error is probabilistically bounded by 
$$\Prob\left\{\|\bm{\mu}(\bm{\rho} \mid \bm{p}, \D_m)-\bm{\rho}(\bm{p})\| \leq\left\|\bm{\beta}^{\top} \Sigma^{\frac{1}{2}}(\bm{\rho} \mid \bm{p}, \D_m)\right\|\right\} \geq \epsilon$$ for $\bm{p} \in \Omega\subset\R^{2d|\mathcal{N}|}$ compact, with  $\epsilon \in(0,1), \bm{\beta}, \bm{\gamma} \in \R^{d}$, and denoting by $m$ the number of entries in the data set $\D_m$, \begin{align}
    \beta_{j} =\sqrt{2\left\|\rho_{j}\right\|_{k}^{2}+300 \gamma_{j} \ln ^{3}\left(\frac{m+1}{1-\epsilon^{1/(d|\mathcal{N}|)}}\right)} 
\end{align}

The variable $\gamma_{j} \in \R$ is the maximum information gain
\begin{align}
    \gamma_{j} &=\max _{\bm{p}^{\{1\}}, \ldots, \bm{p}^{\{m+1\}} \in \Omega} \frac{1}{2} \log \left|I+\sigma_{j}^{-2} K\left(\x, \x^{\prime}\right)\right| \\
    \x, \x^{\prime} & \in\left\{\bm{p}^{\{1\}}, \ldots, \bm{p}^{\{m+1\}}\right\}.
\end{align}
\end{lemma}

\begin{remark}
An efficient algorithm can be used to find $\bm{\beta}$ based on the maximum information gain~\cite{srinivas2012information}.
\end{remark}

\section{Online learning for stable flocking control}\label{sec4}
Consider the potential function $V:\R^{2d|\mathcal{N}|}\to\R$ as described in Section \ref{sec3}.
 In the absence of unknown disturbances, $V$ allows to write the closed-loop system  \cref{formation} as 
\begin{align}
	\begin{cases}
	\dot{\bm{q}} = \nabla_\bm{v} V \\
	\dot{\bm{v}} = -\nabla_\bm{v} V -\nabla_\bm{q} V,
	\end{cases}
	\label{eq: H}
\end{align}
Local exponential convergence to the set \begin{equation}\label{S}\mathcal{S}=\{(\bm{q}^{*},\bm{\delta}^{*})\in\mathbb{R}^{2d|\mathcal{N}|}|\nabla_{\bm{q}}V(\bm{q}^{*})=\bm{0},\bm{\delta}^{*}=\bm{0}\}\end{equation} for the system \cref{eq: H}, has been shown in  \cite{anderson2, sun2015rigid}. Next, we design a decentralized data-driven control law by using GP's, such that, by learning and update the learning of the unknown disturbances, exponentially stabilizes the partially unknown motion of the agents \eqref{double} to a desired formation shape with flocking motion. We will proceed as in \cite{ahn14} for formation control without flocking behaviour, by analyzing an equivalent decoupled gradient systems, by employing a result from \cite{dorfler2011critical}, that we will also use for the design of the decentralized learning-based control law, described as follow. 

Consider the one-parameter family of systems with double integrator flocking stabilization dynamics $\mathcal{H}_\lambda$ given by
\begin{align}
\begin{bmatrix}\dot{\bm{p}} \\ \dot{\bm{v}}\end{bmatrix} = 
\begin{bmatrix}-\lambda I_{d|\mathcal{N}|} & (1-\lambda)I_{d|\mathcal{N}|} \\
	(\lambda-1)I_{d|\mathcal{N}|} & -\mathcal{L}I_{d|\mathcal{N}|}
\end{bmatrix}\begin{bmatrix}\nabla_\bm{p} V \\ \nabla_\bm{v} V \end{bmatrix},
	\label{eq: Hl}
\end{align}
where $\lambda \in [0, 1]$. Equation \cref{eq: Hl} continuously interpolates all convex combinations between the dissipative system (\ref{eq: H}) for $\lambda=0$ and a gradient system for $\lambda=1$. The family $\mathcal{H}_\lambda$ has two important properties summarized in the following Lemma from \cite{dorfler2011critical}.

\begin{lemma}\label{lem: H}
$(I)$ For all $\lambda \in [0, 1]$, the equilibrium set of $\mathcal{H}_\lambda$ is given by the set of the critical points of the potential function $V$, and is independent of $\lambda$. $(II)$ For any equilibrium of $\mathcal{H}_\lambda$ for all $\lambda \in [0, 1]$, the numbers of the stable, neutral, and unstable eigenvalues of the Jacobian of $\mathcal{H}_\lambda$ are the same and independent of~$\lambda$.
\end{lemma}

Denote by $\mathcal{E}_{\bm{e},\bm{\delta}}:=(\bm{e},\bm{\delta})$ the stacked vector of relative positions errors and velocities disagreement vector for stabilization to desired formation shapes with flocking motion. The next theorem introduces the learning-based control law with guaranteed probabilistic boundedness of the error for stabilization under flocking motion.

\begin{theorem}\label{thmain}
Consider the system of agents~\cref{double2} with unknown dynamics and GP models with data sets~\cref{for:dataset} satisfying~\cref{ass:1,ass:2}. Assume that the desired equilibrium set $\mathcal{S}$ given by \cref{S} satisfies~\cref{ass:0}. Then, the control law \begin{align}\label{control-law}
    \bm{u}(t)=-\mathcal{L}\bm{v}-R^\top(\bm{e})\bm{e}(\bm{z})-\hat{\f}(\bm{q},\bm{v})-\bm{\mu}(\bm{\rho}|\bm{p},\D_m)
\end{align} guarantees that the solution trajectories converge locally exponentially fast to the equilibrium set $\mathcal{S}$ and are ultimately uniformly bounded in probability by \begin{align}
    \Prob\{||\mathcal{E}_{\bm{e},\bm{\delta}}(t)||\leq\sqrt{2}\max_{p\in\Omega}\bar{\Delta}_{m(T)}(\bm{p}),\forall t\geq T_\epsilon\}\geq \epsilon
\end{align} on $\Omega$ with $T_\epsilon\in\R_{\geq 0}$. 
\end{theorem}

\begin{remark}Note that the individual control law $\bm{u}_i(t)$ of each agent depends on the distance and velocity to its neighbors and the data set based on its own dynamics only. \end{remark}

\textit{Proof}: Note that the relative position error \eqref{e_k} satisfies $\dot{\bm{e}}_k=2\bm{z}_k^{T}\cdot\bm{z}_k=2D^{T}(\bm{z})\bar{B}(\bm{z})\bm{v}$, that is, $\dot{\bm{e}}=2R(\bm{z})\bm{v}$. Using that $\dot{\bm{v}}=\bm{\dot{\delta}}$, the evolution for the relative position error can be written in terms of the rigidity matrix and the velocity disagreement vector as $\dot{\bm{e}}=2R(\bm{z})\bm{\delta}$.

Denoting by $\mathcal{E}_{\bm{e},\bm{\delta}}^{\lambda}$ the stacked vector of errors $\mathcal{E}_{\bm{e},\bm{\delta}}$ from~\cref{eq: Hl} for any $\lambda\in[0,1]$, which includes the closed-loop system~\cref{formation} for $\lambda=1$, we know that as a consequence of Lemma \ref{lem: H}, $\mathcal{E}_{\bm{e},\bm{\delta}}^{\lambda}$ and $\mathcal{E}_{\bm{e},\bm{\delta}}$ share the same stability properties. By using Lemma \ref{lem: H}, we will study the system (\ref{eq: Hl}) for $\lambda = 0.5$, without loss of generality, that is, \begin{align}
	\dot{\bm{p}} &= -\frac{1}{2}\overline BD(\bm{z})\bm{e} + \frac{1}{2}\bm{\delta} \label{eq:p} \\
	\dot{\bm{z}} &= -\frac{1}{2}\overline B^T\overline BD(\bm{z})\bm{e} + \frac{1}{2}\overline B^T \bm{\delta} \label{eq: zl} \\
	\dot{\bm{e}} &= -D(\bm{z})^T\overline B^T\overline BD(\bm{z})\bm{e} + D(\bm{z})^T\overline B^T \bm{\delta} \label{eq: el} \\
	\dot{\bm{\delta}} &= -\frac{1}{2}\overline BD(\bm{z})\bm{e} - \mathcal{L}\bm{\delta} \label{eq: vl}.
\end{align}
Consider the Lyapunov candidate function $V$ for the system \cref{double2} with control law $\bm{u}(t)$ as in \cref{control-law}, given by the positive definite and radially unbounded function 
\begin{align}\label{V}
V(\bm{e},\bm{\delta})=\frac{1}{2}||\bm{e}||^2+||\bm{\delta}||^2.
\end{align}  Note that the time-derivative of the Lyapunov function along the closed-loop trajectory satisfies
\begin{align}
    \dot{V}&=\bm{e}^\top\dot{\bm{e}}+2\bm{\delta}^\top\dot{\bm{\delta}}\\
    &=-\begin{bmatrix}\bm{e} & \bm{\delta} \end{bmatrix}\begin{bmatrix} D(\bm{z})^\top\overline B^T\overline BD(\bm{z}) & D(\bm{z})^\top\overline B^\top\\ \overline BD(\bm{z}) &  \mathcal{L}\end{bmatrix}\begin{bmatrix}\bm{e} \\ \bm{\delta} \end{bmatrix}\notag\\
    &+\bm{\delta}^\top(\bm{\rho}(\bm{p})-\bm{\mu}(\bm{\rho}|\bm{p},\D_m))\notag\\
    &=-\bm{e}^\top R(\bm{z})R(\bm{z})^\top\bm{e}-\bm{\delta}^\top\mathcal{L}\bm{\delta}+\bm{\delta}^\top(\bm{\rho}(\bm{p})-\bm{\mu}(\bm{\rho}|\bm{p},\D_m)).\notag
\end{align}

Denote by $\lambda_{min}$ and $\lambda_{2}$ the minimum eigenvalue of $R(\bm{z})R^\top(\bm{z})$ and the non-zero minimum eigenvalue of $\mathcal{L}$, respectively. By ~\cref{ass:0} the rigidity matrix is full rank (except the non-generic cases,  e.g., collinear or coplanar alignments of the agents in $\R^2$ or $\R^3$). Therefore $\lambda_{min}>0$. Note also that, by ~\cref{ass:0},  $\lambda_{2}>0$. Therefore, by employing Lemma \ref{model-error} it follows that
\begin{align}\label{for:Lyapevo}
  \Prob\{\dot{V}\leq-\lambda_{\min}||\bm{e}||^2-\lambda_{2}||\bm{\delta}||^2+||\bm{\delta}||\bar{\Delta}_m(\bm{p})\}\geq\epsilon,  
\end{align}
where $\bar{\Delta}_m(\bm{p}):\Omega\to\R_{\geq 0}$ is a bounded function such that $\Vert\bm{\beta}^{\top} \Sigma^{\frac{1}{2}}(\bm{\rho}\mid \bm{p},\D_m)\Vert\leq\bar{\Delta}_m(\bm{p})$, which exists because the kernel function is continuous and therefore it is bounded on a compact set $\Omega\subset\R^{2d|\mathcal{N}|}$, and then the variance $\Sigma(\bm{\rho}\mid \bm{p},\D_m)$ is bounded, see \cite{beckers2016equilibrium}. Then, the value of $\dot{V}$ is negative with probability $\epsilon$ for all $\mathcal{E}_{\bm{e},\bm{\delta}}$ with $\displaystyle{||\mathcal{E}_{\bm{e},\bm{\delta}}||>\max_{\bm{p}\in\Omega}\sqrt{2}\bar{\Delta}_m(\bm{p})}$, where the maximum exists since $\bar{\Delta}_m(\bm{p})$ is bounded in $\Omega$. By using~\cref{ass:1}, we define $T\in\R_{\geq 0}$ such that $\D_{m(T)}=\D_{m(t)}$ for all $t\geq T$. Then, $V$ is uniformly ultimately bounded in probability by $\displaystyle{\Prob\{||\mathcal{E}_{\bm{e},\bm{\delta}}||\leq b,
\,\forall t\geq T_\epsilon\in\R_{\geq 0}\}\geq\epsilon}$ with probabilistic bound $\displaystyle{b=\max_{\bm{p}\in\Omega}\sqrt{2}\bar{\Delta}_{m(T)}(\bm{p})}$.\hfill$\square$

\begin{remark}
Regions of attraction for $\lambda=\frac{1}{2}$ might be different from $\lambda= 1$ since Lemma \ref{lem: H} depends of the linearization around the equilibrium (i.e., the Jacobian for \eqref{eq: Hl}). Note also that finding explicitly a probabilistic ultimate bound required to solve an optimization problem on $\mathbb{R}^{2d|\mathcal{N}|}$ constrained to the compact set $\Omega$. 
\end{remark}

\section{Simulation results}\label{sec5}
Next, we present three numerical simulations to illustrate the performance of the control law proposed in Theorem \ref{thmain} with different communication graphs. 

\subsection{3 Agents in 2d Space}
\begin{figure}[h]
	\begin{tikzpicture}
	\node at (0,0) {\includegraphics[width=8cm]{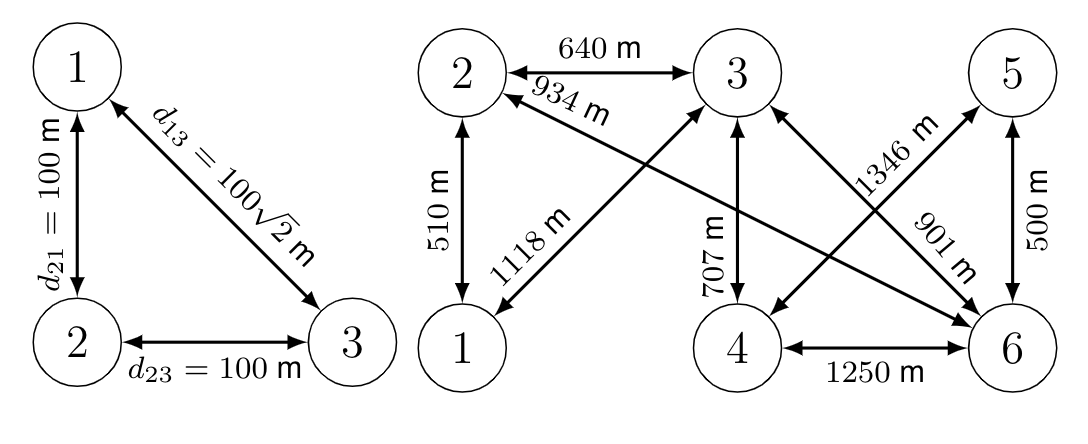}};
	\draw[thick,dashed] (-1,-1.6) -- (-1,1.6);
	\node at (-2.3,-1.5) {Scenario 1};
	\node at (1.5,-1.5) {Scenario 2};
	\end{tikzpicture}
	\caption{Neighbor’s relations and desired shapes for two different scenarios.}
	\label{fig:graph}
\end{figure}
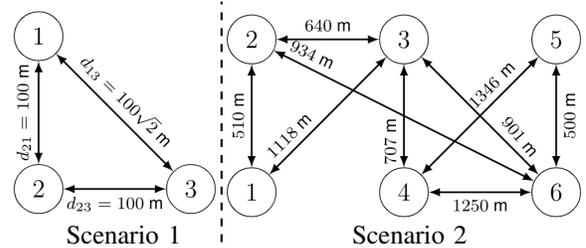

The first scenario consists of $n=3$ planar agents (i.e., $d=2$). The position of each agent $i\in\mathcal{N}$ is denoted by $\bm{q}_i=[x_i,y_i]^\top$. The neighbor’s relations and the desired shape are depicted on the left in~\cref{fig:graph}.

By introducing a perturbation over the velocities of the agents, we can compare the nominal control law proposed in \cite{anderson2} and \cite{sun2015rigid} with the data-driven control law \eqref{control-law}. Different perturbations had been chosen on purpose in order to highlight the \textit{losses in cohesion, alignment and separation}. In order to better observe this behaviour, we also provide a video for the numerical experiment\footnote{https://youtu.be/X8-uaqzPqtc}.

For losses in \textit{cohesion}, assume agents 1 and 3 are affected by the unknown dynamics given by $\f_1(\bm{v}_1)=[-300\sin(0.01 v_{1,y})-50, -300]^\top\notag$, $\f_3(\bm{v}_3)=[300\sin(0.01 v_{3,y}), 300]^\top$. These external forces undermine the cohesion property in the motion, causing agents to separate themselves. Respectively, losses in \textit{alignment} are observed under a perturbation over the first agent given by $ \f_1(\bm{v}_1)=[-300\sin(0.01 v_{1,y})-100, -300\sin(0.01 v_{1,x})-100]^\top\notag$. 

\begin{figure}[htbp!]
    \centering
    \begin{subfigure}[b]{0.5\linewidth}
    \centering
    \includegraphics[width=\linewidth]{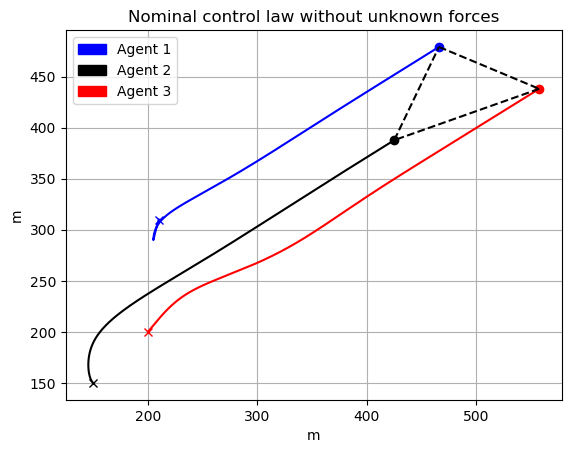} 
    \caption{Nominal control law with\\ known model}   \label{2fig7:a} 
    \vspace{4ex}
  \end{subfigure}
  \begin{subfigure}[b]{0.5\linewidth}
    \centering
   \includegraphics[width=\linewidth]{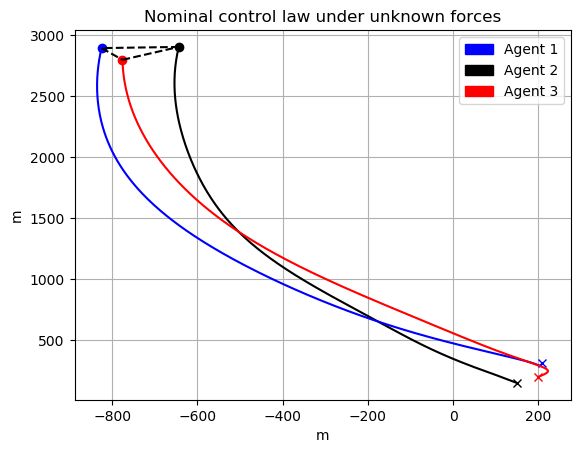}
    \caption{Nominal control with\\ unknown model} 
    \label{2fig7:b} 
    \vspace{4ex}
  \end{subfigure} 
  \caption{Undesired flocking motion of the 3 agents with the nominal control law}
  \label{2Pert}
\end{figure}

\begin{figure}[htbp]
    \centering
    \begin{subfigure}[b]{0.5\linewidth}
    \centering
    \includegraphics[width=\linewidth]{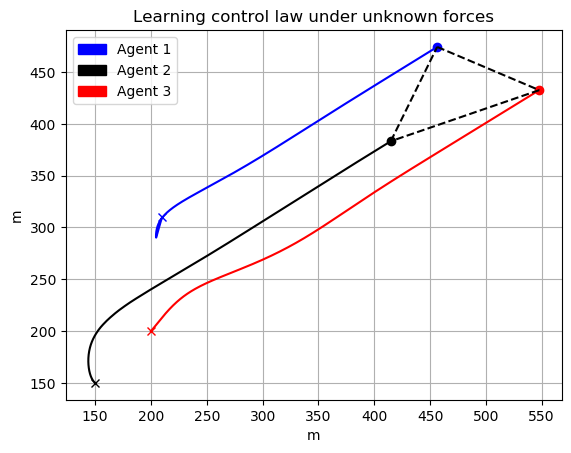} 
    \caption{Stable flocking with\\ online learning}   \label{2stable}
    \vspace{4ex}
  \end{subfigure}
  \begin{subfigure}[b]{0.5\linewidth}
    \centering
   \includegraphics[width=\linewidth]{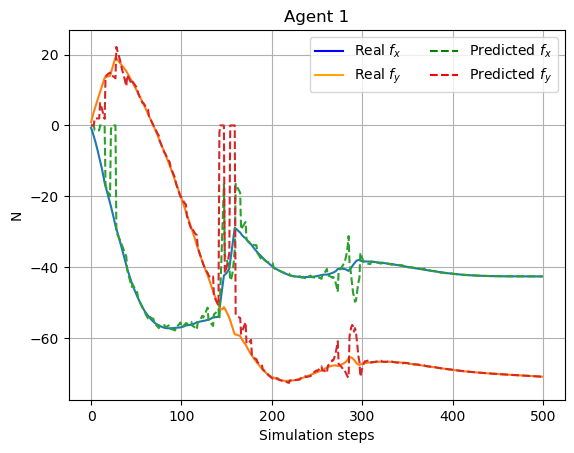}
    \caption{Agent $1$: real/predicted\\ external force}
  \label{2ag4}
    \vspace{4ex}
  \end{subfigure} 
 \caption{Stable flocking with learning-based control for 3 agents and the real/predicted forces for the first agent}\label{2stablebig}
\end{figure}

The simulation video\footnotemark[1] shows the loss of alignment in flocking motion under $\f_1(\bm{v}_1)$. This means that each agent's speed is no longer synchronize with that of their neighbors, causing undesired motion behaviours. By employing Theorem \ref{thmain}, the effect of $\f_1$ is compensated by the data-driven control law \eqref{control-law} and the agents can keep a constant speed while preserving the desired shape. In addition, in Figure \ref{2Pert} we show the implementation of the nominal control law (Fig. \ref{2fig7:a}) and the controlled system under the influence of $\f_1(\bm{v}_1)=[-200\sin(0.01 v_{1,y}), -200\sin(0.01 v_{1,X})]^\top\notag$ (Fig. \ref{2fig7:b}). Figure \ref{2stablebig} shows the implementation of the data-driven control law (Fig. \ref{2stable}) and the comparison between the unknown dynamics affecting each agent's motion and the predicted motions based on the online learning (Fig. \ref{2ag4}). 

An undesired \textit{separation} behavior might occur when agents $2$ and $3$ are perturbed under $\f_2(\bm{v}_1)=[-30\sin(0.01 v_{1,y}), 100\sin(0.01 v_{1,x})+50]^\top\notag$ and $\f_3(\bm{v}_3)=[-300\sin(0.01 v_{3,x})-200, 300]^\top$, so there is a collision between agents. As we show in the video\footnotemark[1], by following Theorem \ref{thmain}, the collision is successfully avoided.
\subsection{6 Agents in 2d Space}
\begin{figure}[htbp]
    \centering
    \begin{subfigure}[b]{0.5\linewidth}
    \centering
    \includegraphics[width=\linewidth]{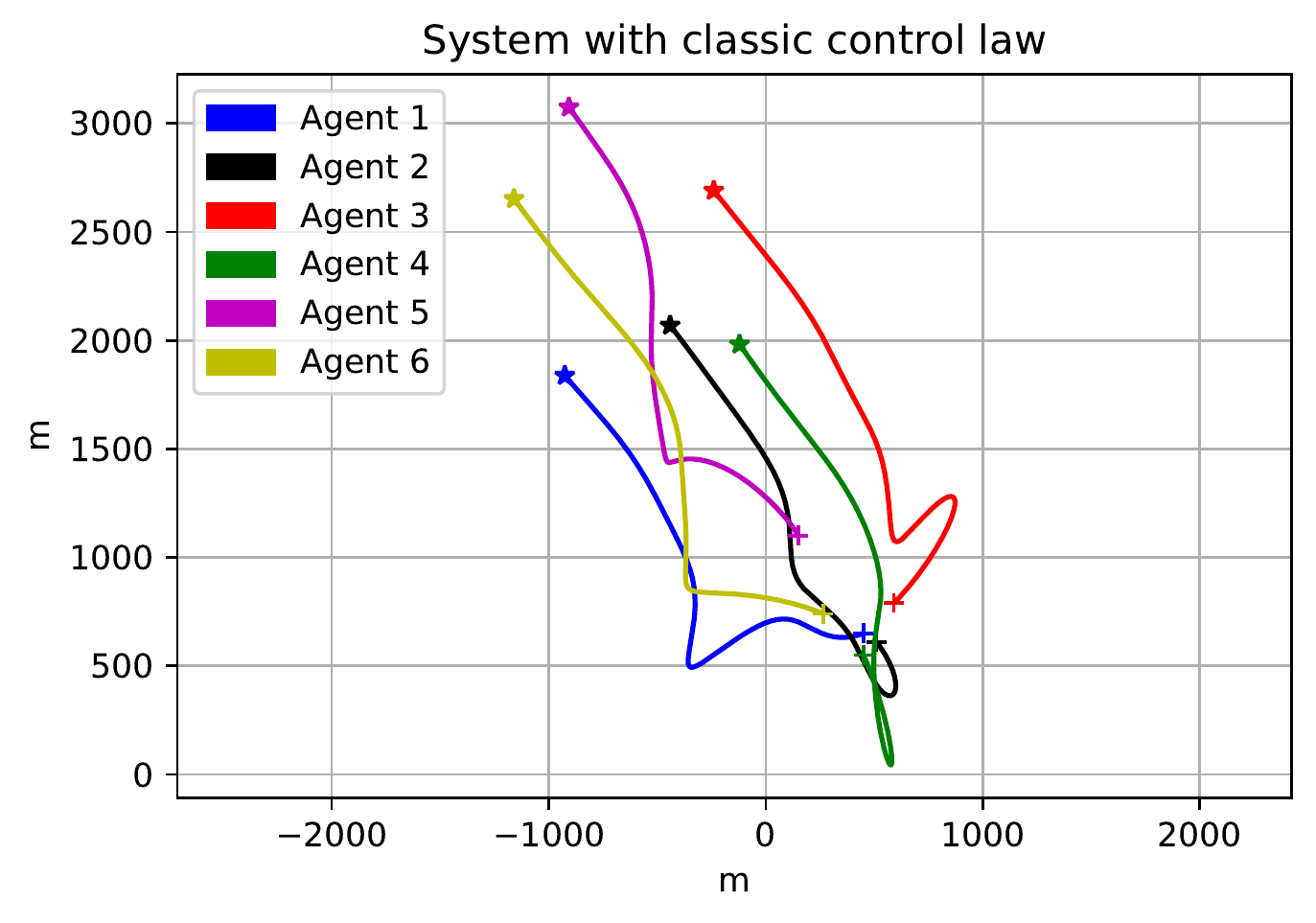} 
    \caption{Nominal control law with\\ known model}   \label{fig7:a} 
    \vspace{4ex}
  \end{subfigure}
  \begin{subfigure}[b]{0.5\linewidth}
    \centering
   \includegraphics[width=\linewidth]{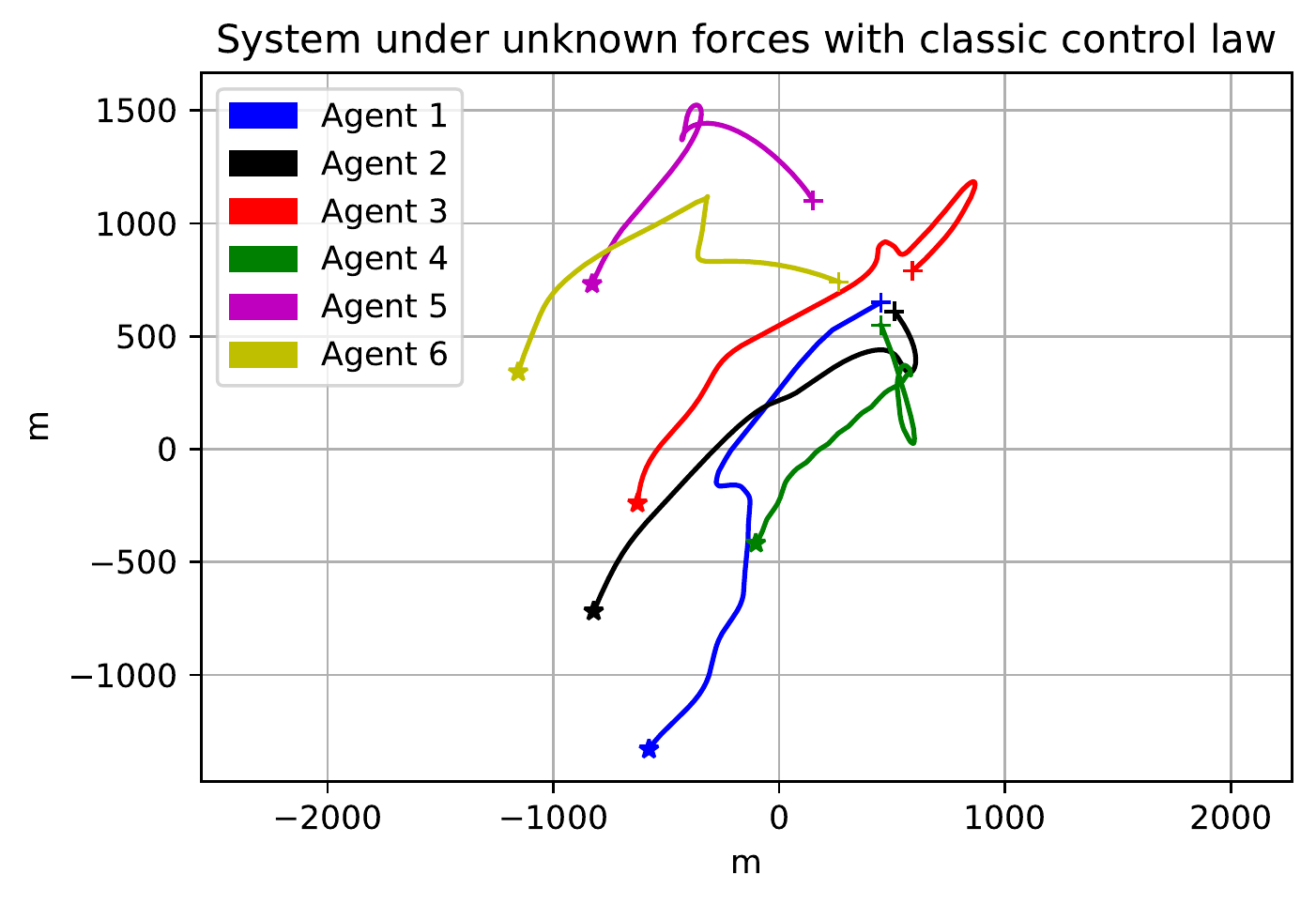}
    \caption{Nominal control with\\ unknown model} 
    \label{fig7:b} 
    \vspace{4ex}
  \end{subfigure} 
  \caption{Undesired flocking motion of 6 agents with the nominal control law}
  \label{Pert}
\end{figure}

\begin{figure}[htbp]
    \centering
    \begin{subfigure}[b]{0.5\linewidth}
    \centering
    \includegraphics[width=\linewidth]{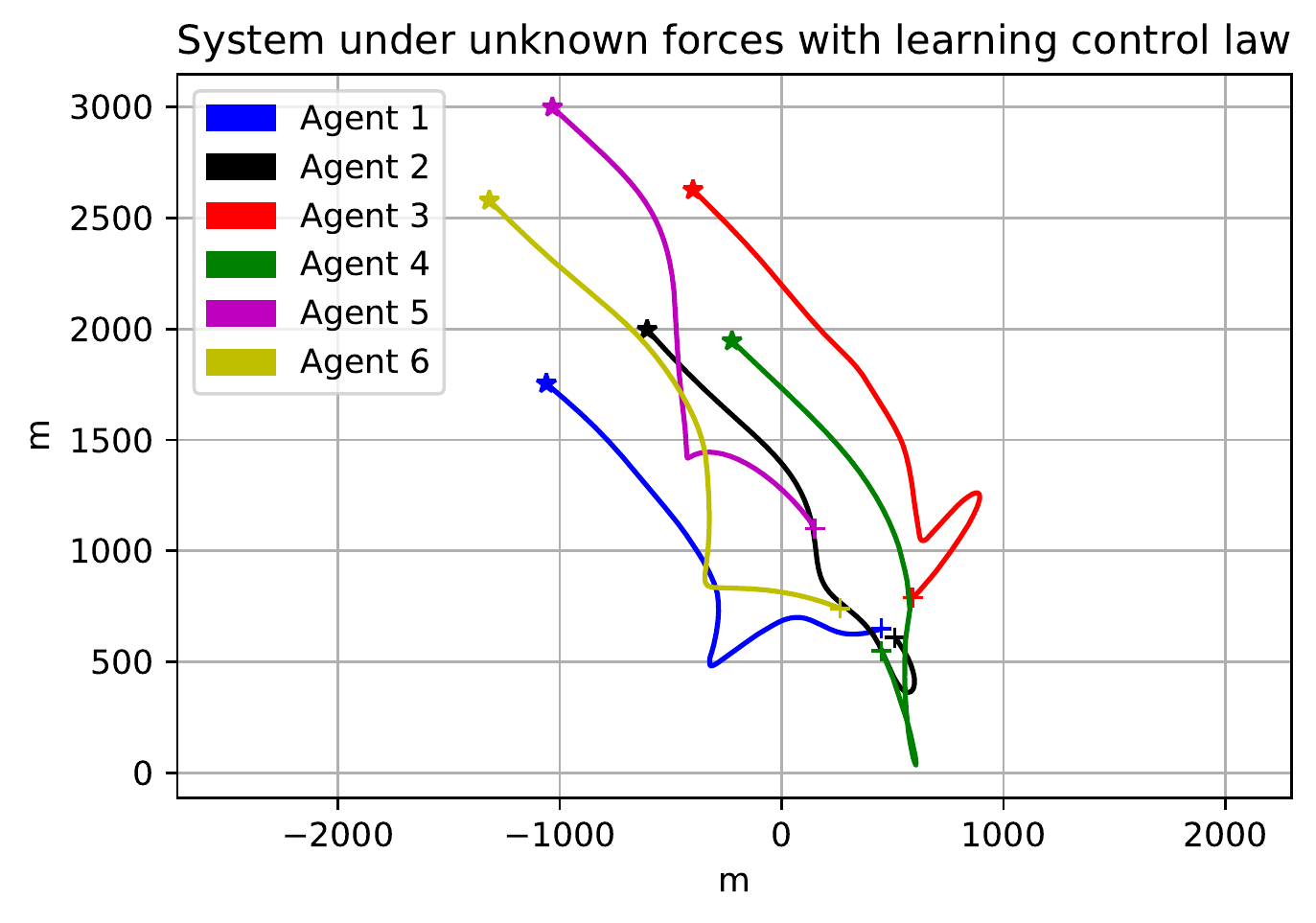} 
    \caption{Stable flocking with\\online learning}   \label{stable}
    \vspace{4ex}
  \end{subfigure}
  \begin{subfigure}[b]{0.5\linewidth}
    \centering
   \includegraphics[width=\linewidth]{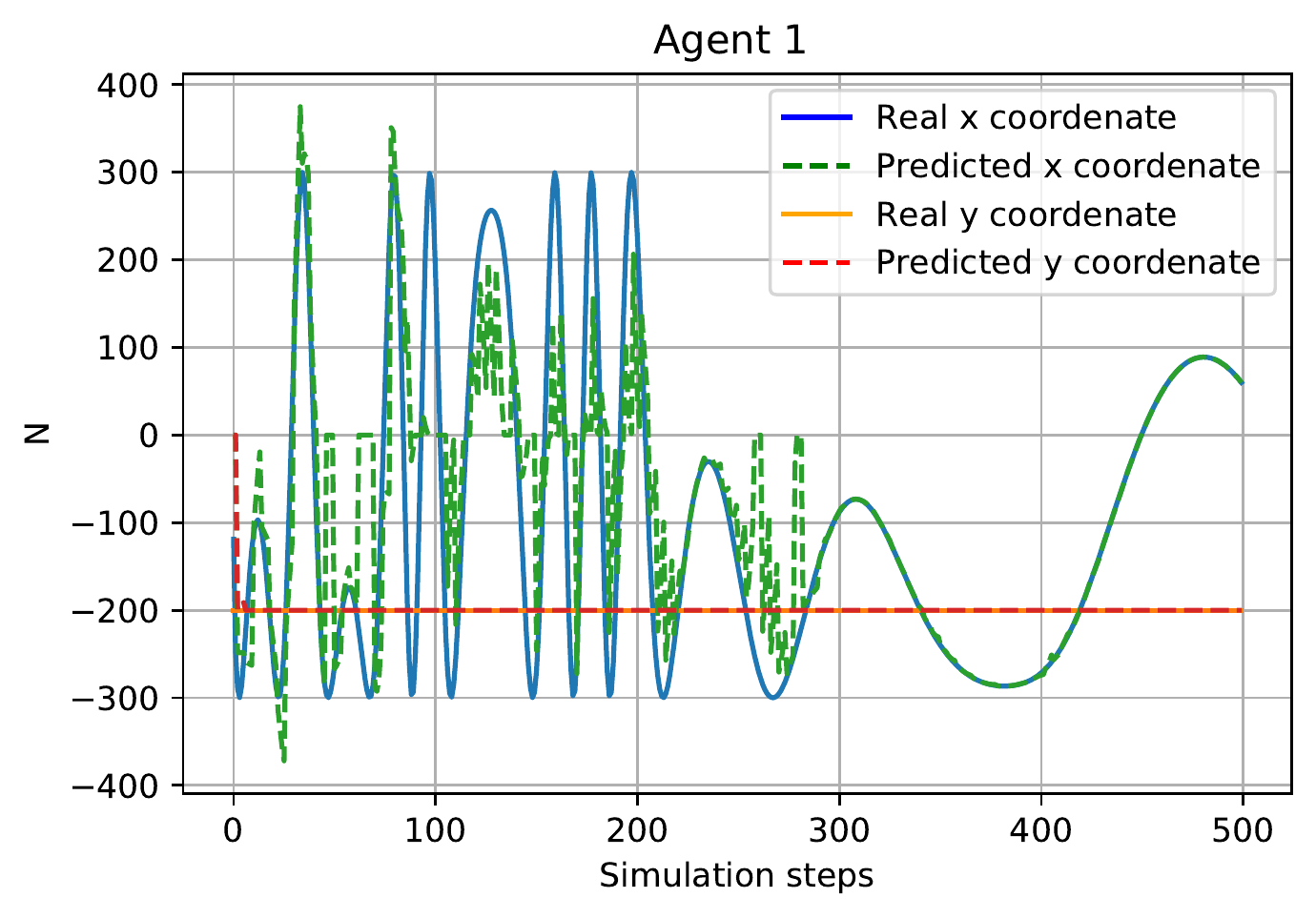}
    \caption{Agent $1$: real/predicted\\ external force}
  \label{ag4}
    \vspace{4ex}
  \end{subfigure} 
 \caption{Stable flocking with learning-based control for 6 agents and the real/predicted forces for the first agent}\label{stablebig}
\end{figure}

We assume agents $1$, $3$ and $4$ are perturbed by unknown external forces $\f_{1}$, $\f_{3}$ and $\f_{4}$, respectively, given by $\f_1(\bm{v}_1)=[300\sin(0.2 v_{1,y}), -200]^\top$, $\f_3(\bm{v}_3)=[300\sin(0.2 v_{3,y}), -200]^\top$ and 
    $\f_4(\bm{v}_4)=[-300\sin(0.2 v_{4,y}), 300\cos(0.2 v_{4,x})]^\top$. Under initial positions $\bm{q}_i(0)$ given by $\bm{q}_1(0)=(450,200)$, $\bm{q}_2(0)=(510,100)$, $\bm{q}_3(0)=(590,300)$, $\bm{q}_4(0)=(450,0)$, $\bm{q}_5(0)=(250,650)$, $\bm{q}_6(0)=(265,400)$, in Figure \ref{Pert}, we observe the effects of perturbations with the nominal control law. Agents positions are skewed down due to the unknown forces. However, the data-driven control law \eqref{control-law} manages to compensate this tendency, as shown in Figure \ref{stable}. 

\begin{figure}[htbp]
    \centering
    \begin{subfigure}[b]{0.5\linewidth}
    \centering
    \includegraphics[width=\linewidth]{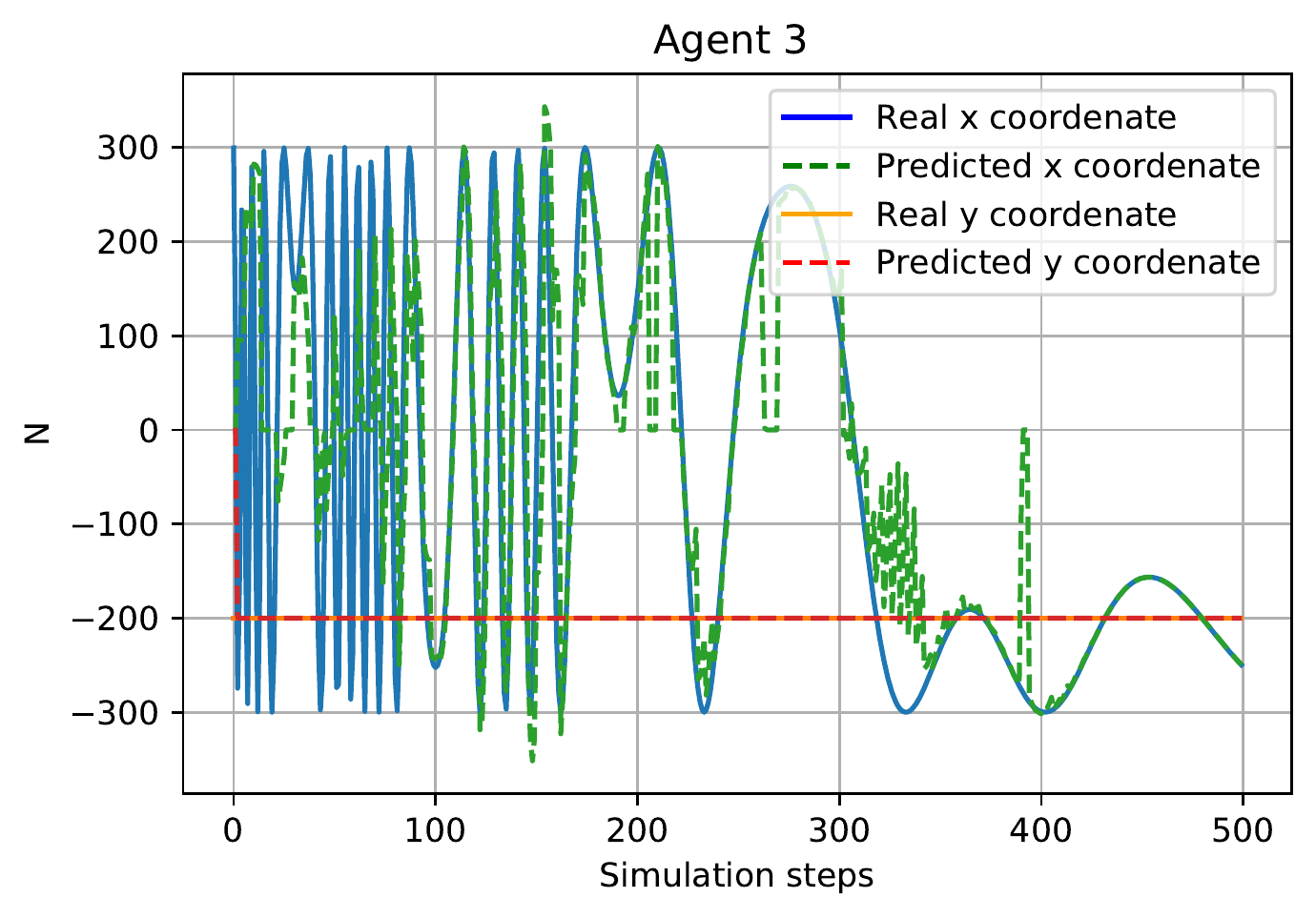} 
    \caption{Agent $3$: real/predicted\\ external force}   \label{Pert2}
    \vspace{4ex}
  \end{subfigure}
  \begin{subfigure}[b]{0.5\linewidth}
    \centering
   \includegraphics[width=\linewidth]{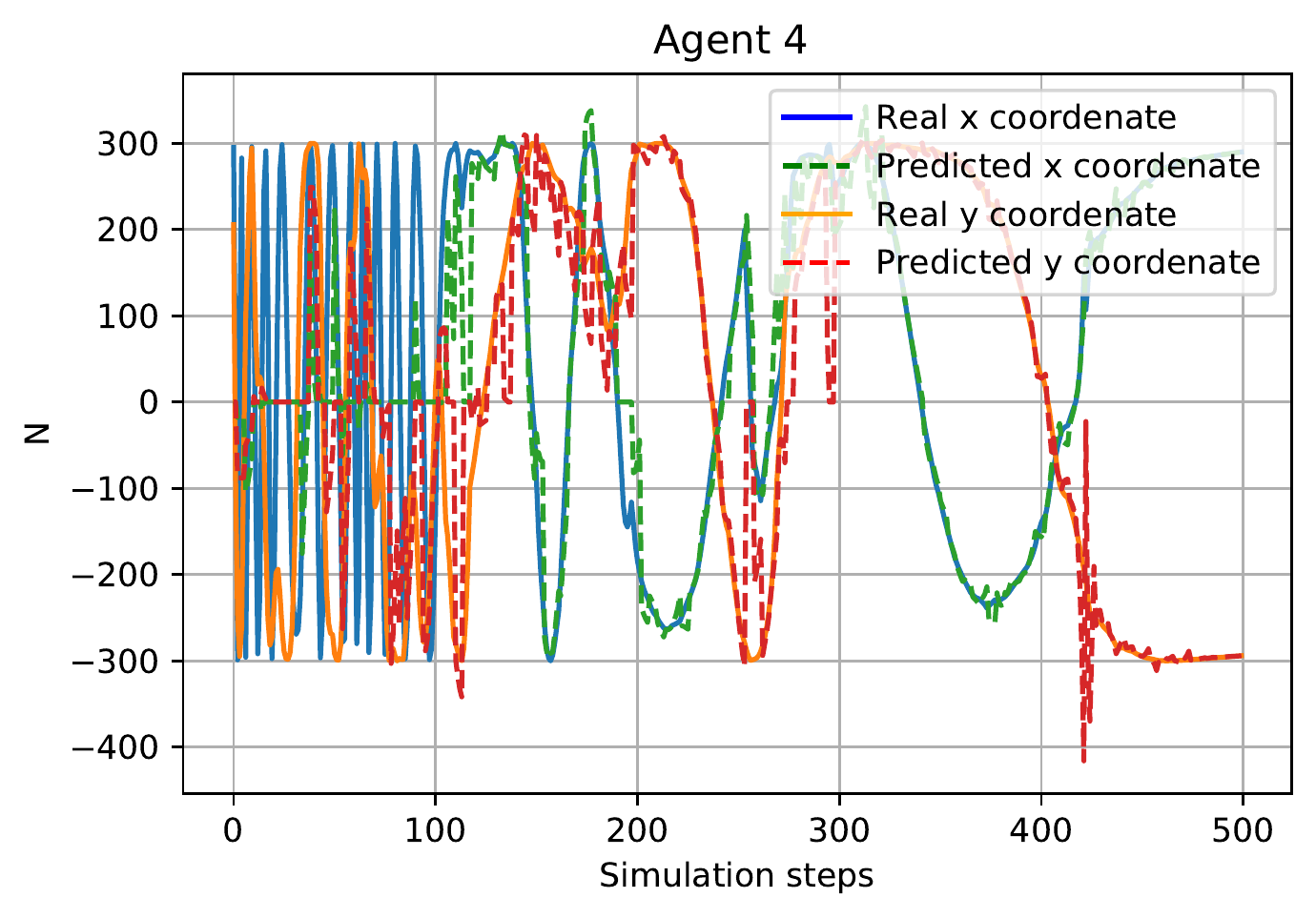}
    \caption{Agent $4$: real/predicted\\ external force}
  \label{ag4rp}
    \vspace{4ex}
  \end{subfigure} 
  \vspace{-0.5cm}
 \caption{Real/predicted external force for Agents 3 and 4}\label{ag34}
\end{figure}

\begin{figure}[htbp]
    \centering
    \begin{subfigure}[b]{0.5\linewidth}
    \centering
    \includegraphics[width=\linewidth]{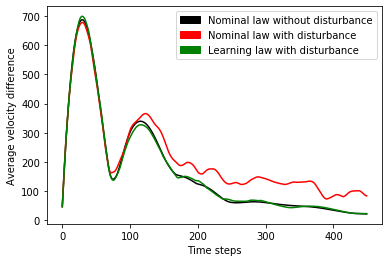} 
    \caption{Average velocity}   \label{average}
    \vspace{4ex}
  \end{subfigure}
  \begin{subfigure}[b]{0.5\linewidth}
    \centering
   \includegraphics[width=\linewidth]{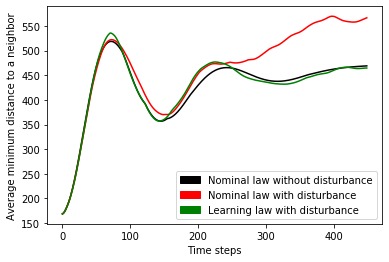}
    \caption{Average distance}
  \label{distance}
    \vspace{4ex}
  \end{subfigure} 
    \vspace{-0.3cm}
 \caption{Evaluation of the performance with different metrics}\label{metrics}
\end{figure}

GPR's prediction capability can be observed in Figures \ref{ag4} and \ref{ag34} by means of a comparison between the unknown dynamics affecting each agent's motion and the predicted motions based on the online learning. The first two Figures show how the same external force has different behaviours over agents $1$ and $3$, due to the different role they play in the overall motion of the system, while Figure \ref{ag4rp} shows the unknown force introduced is different and although the $y$-coordinate is not constant, the proposed learning-based control law does well managing both components. 
In addition, inspired by \cite{Ji2021}, we evaluate the performance with some metrics. In particular, we study the average velocity to provide a measure of the alignment of the velocities of the agents. We observe in Figure \ref{average} how well the velocities of agents are aligned. In addition, we measure the average distance among nearest neighbors. As it is expected in distance-based formation control, Figure \ref{distance} show that agents stabilize their distances once they reach the desired shape.

\subsection{4 Agents in 3d Space}
\begin{figure}[h]
\centering
       \includegraphics[width=3.5cm]{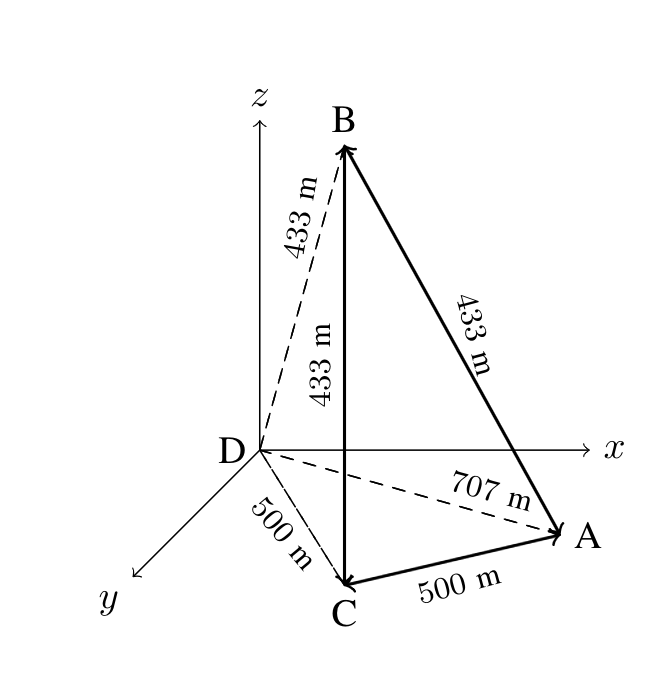}
	\caption{Neighbor’s relations and desired shape}
	\label{fig:graphtetra}
\end{figure}

Finally, we show the applicability of our method for learning flocking control of agents moving on the $3$-dimensional space. We consider $4$ agents forming a shape as depicted in Figure \ref{fig:graphtetra}.

\begin{figure}[htbp]
    \centering
    \begin{subfigure}[b]{0.5\linewidth}
    \centering
    \includegraphics[width=\linewidth]{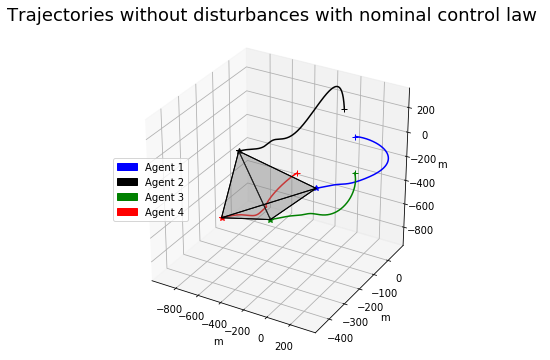} 
    \caption{Nominal control law in\\ known model}   \label{fig7:a2} 
    \vspace{4ex}
  \end{subfigure}
  \begin{subfigure}[b]{0.47\linewidth}
    \centering
   \includegraphics[width=\linewidth]{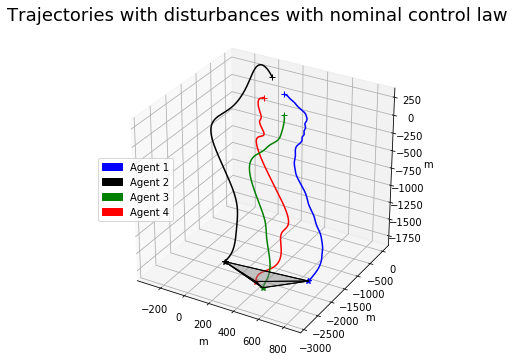}
    \caption{Nominal control in\\ unknown model} 
    \label{fig7:b2} 
    \vspace{4ex}
  \end{subfigure} 
  \vspace{-0.3cm}
  \caption{Undesired flocking motion with nominal control law}
  \label{3dnom}
\end{figure}

Consider the perturbations over agents $1$ and $3$ given by $\f_1(\bm{v}_1)=[300\sin(0.2 v_{1,y}),300\cos(0.2 v_{1,x}) ,10]^\top$, $\f_3(\bm{v}_3)=[300\sin(0.2 v_{3,y}), -200,300\sin(0.2 v_{3,y})]^\top$. Initial positions are
$\bm{q}_1(0)=(100,0,0)$, $\bm{q}_2(0)=(0,0,200)$, $\bm{q}_3(0)=(0,-300,0)$, $\bm{q}_4(0)=(100,0,-300)$.
Note the significant deterioration of the motion when the classical control law has to deal with the unknown forces, as observed in Figure \ref{3dnom}.

\begin{figure}[htbp]
    \centering
    \begin{subfigure}[b]{0.5\linewidth}
    \centering
    \includegraphics[width=\linewidth]{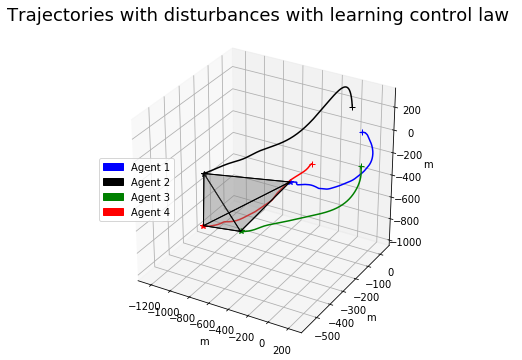} 
    \caption{Nominal control law in known model}   
    \vspace{4ex}
  \end{subfigure}
  \begin{subfigure}[b]{0.47\linewidth}
    \centering
   \includegraphics[width=\linewidth]{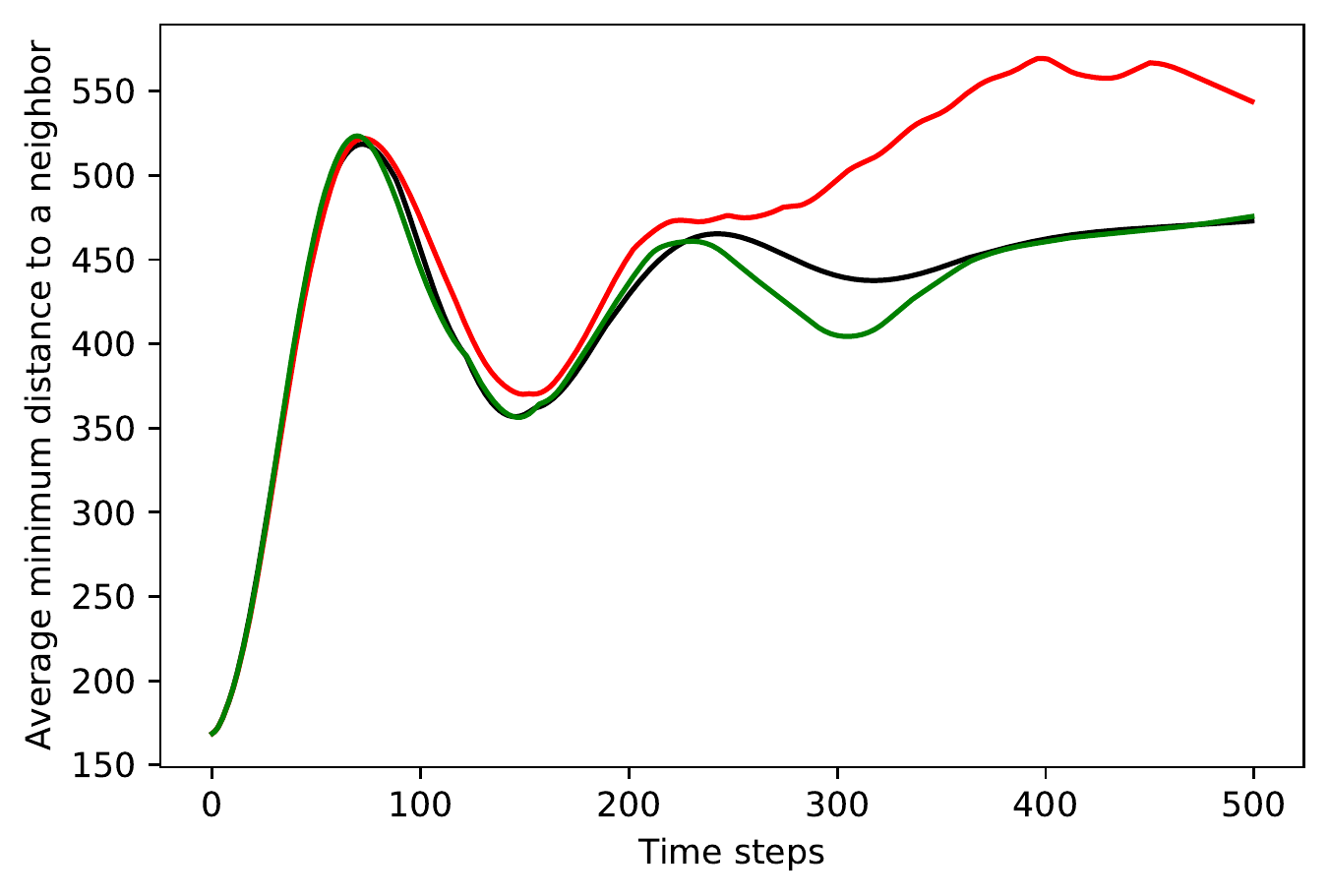}
    \caption{Average distance} 
    \vspace{4ex}
  \end{subfigure} 
    \vspace{-0.3cm}
  \caption{Stable flocking with learning-based control and average distance}
  \label{3dlearn}
\end{figure}

\begin{figure}[htbp]
    \centering
    \begin{subfigure}[b]{0.5\linewidth}
    \centering
    \includegraphics[width=\linewidth]{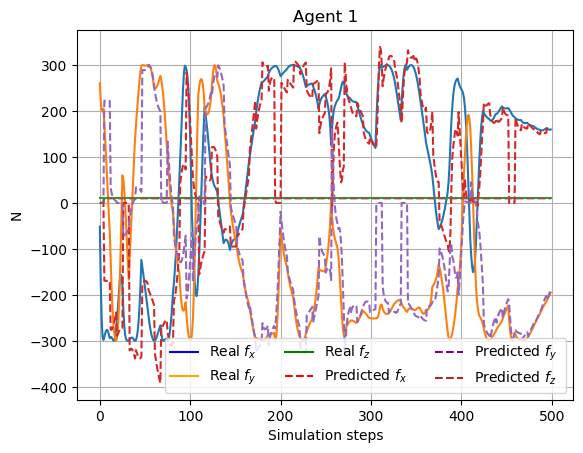} 
    \caption{Agent $1$: real/predicted\\external force}   \label{Pert2b}
    \vspace{4ex}
  \end{subfigure}
  \begin{subfigure}[b]{0.5\linewidth}
    \centering
   \includegraphics[width=\linewidth]{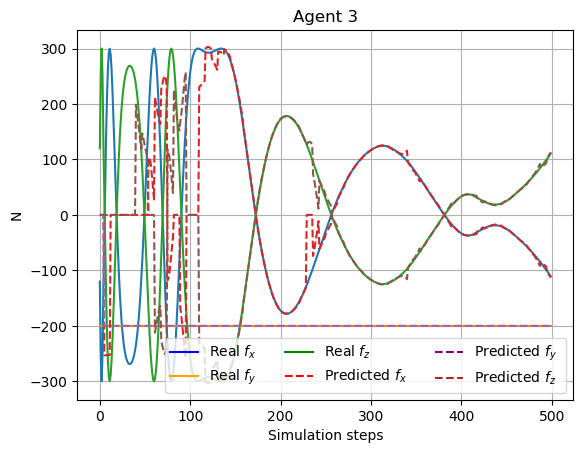}
    \caption{Agent $3$: real/predicted\\ external force}
    \vspace{4ex}
  \end{subfigure}
    \vspace{-0.3cm} 
 \caption{Real/predicted external force for Agents 1 and 3}\label{ag133d}
\end{figure}

As in the $2$D situation, with the learning control law, it can be observed in Figure \ref{3dlearn} how the perturbation produced on the agents motion under the unknown external forces is overcame, as well as the average distance. Figure \ref{ag133d} shows the learning process, respectively, in agents $1$ and $3$, displaying the comparison between the real and predicted external forces. 

\begin{figure}[h]
\centering
       \includegraphics[width=7.5cm]{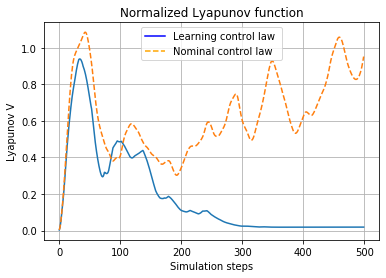}
	\caption{Normalized Lyapunov function with the standard control law (dashed) and the proposed, learning-based control law (solid) which converges to a tight set around zero.}
	\label{fig:V}
\end{figure}

Finally, the evolution of the Lyapunov function in~\cref{fig:V} highlights the superior of the proposed control law as it allows the Lyapunov function to converge to a tight set around zero. Note that the evolution of the Lyapunov function is not always decreasing but bounded in a neighborhood around zero, see~\cref{for:Lyapevo}. The size of the set shrinks with improved accuracy of the GP model.

\section{Conclusions}
We present an online learning-based control law for stable flocking of second-order agents with unknown dynamics. By employing Gaussian Processes, the error to achieve desired flocking motions is proven to be bounded in probability. The online framework allows to improve the quality of the GPs over time and, thus, to improve stability performance. Finally, numerical examples visualize the effectiveness of the proposed control law.


\section*{Acknowledgements}

This work was supported by a $2020$ Leonardo Grant for Researchers and Cultural Creators, BBVA Foundation. M. G. and L. C. are also supported by MICINN grant PID2019-106715GB-C21, ``Severo Ochoa Programme for Centres of Excellence in R$\&$D''
(CEX$2019$-$000904$-$S$) and from the Spanish National Research Council, through the ``Ayuda extraordinaria a Centros de Excelencia Severo Ochoa''(20205-CEX001). The BBVA Foundation accepts no responsibility for the opinions, statements and contents included in the project and/or the results thereof, which are entirely the responsibility of the authors.

\bibliographystyle{IEEEtran}
\bibliography{root}


\end{document}

%% file: mydefs.tex
\newtheorem{assum}{Assumption}

\newcommand\tran{\mkern-2mu\raise1.25ex\hbox{$\scriptscriptstyle\top\hspace{0.5mm}$}\mkern-3.5mu}
\newcommand{\R}{\mathbb{R}}

\newcommand{\D}{\mathcal{D}}

\newcommand{\bm}[1]{{\boldsymbol{#1}}}

\DeclareMathOperator{\diag}{diag}
\DeclareMathOperator{\var}{var}

\DeclareMathOperator{\rank}{rank}

\DeclareMathOperator{\Prob}{P}

\newcommand{\x}{\bm x}

\newcommand{\f}{\bm{f}}

\renewcommand{\u}{\bm{u}}
\newcommand{\y}{\bm{y}}

\usepackage[noabbrev]{cleveref} 
\crefname{rem}{Remark}{Remarks}
\crefname{exam}{Example}{Examples}
\crefname{assum}{Assumption}{Assumptions}
\crefname{prop}{Proposition}{Propositions}
\crefname{propy}{Property}{Properties}
\crefname{cor}{Corollary}{Corollaries}
\crefname{lem}{Lemma}{Lemmas}
\crefname{section}{Section}{Sections}
\crefname{subsection}{Section}{Sections}
\crefname{thm}{Theorem}{Theorems}
\crefname{defn}{Definition}{Definitions}
\crefname{figure}{Fig.}{Fig.}
\Crefname{figure}{Figure}{Figures}
\crefname{equation}{}{}

%% file: root.bbl
\begin{thebibliography}{10}
\providecommand{\url}[1]{#1}
\csname url@samestyle\endcsname
\providecommand{\newblock}{\relax}
\providecommand{\bibinfo}[2]{#2}
\providecommand{\BIBentrySTDinterwordspacing}{\spaceskip=0pt\relax}
\providecommand{\BIBentryALTinterwordstretchfactor}{4}
\providecommand{\BIBentryALTinterwordspacing}{\spaceskip=\fontdimen2\font plus
\BIBentryALTinterwordstretchfactor\fontdimen3\font minus
  \fontdimen4\font\relax}
\providecommand{\BIBforeignlanguage}[2]{{%
\expandafter\ifx\csname l@#1\endcsname\relax
\typeout{** WARNING: IEEEtran.bst: No hyphenation pattern has been}%
\typeout{** loaded for the language `#1'. Using the pattern for}%
\typeout{** the default language instead.}%
\else
\language=\csname l@#1\endcsname
\fi
#2}}
\providecommand{\BIBdecl}{\relax}
\BIBdecl

\bibitem{rubenstein2014}
M.~Rubenstein, A.~Cornejo, and R.~Nagpal, ``Programmable self-assembly in a
  thousand-robot swarm,'' \emph{Science}, vol. 345, no. 6198, pp. 795--799,
  2014.

\bibitem{reynolds1987flocks}
C.~W. Reynolds, ``Flocks, herds and schools: A distributed behavioral model,''
  in \emph{Proceedings of the 14th annual conference on Computer graphics and
  interactive techniques}, 1987, pp. 25--34.

\bibitem{tanner2003stable}
H.~G. Tanner, A.~Jadbabaie, and G.~J. Pappas, ``Stable flocking of mobile
  agents, part i: Fixed topology,'' in \emph{42nd IEEE International Conference
  on Decision and Control}, vol.~2.\hskip 1em plus 0.5em minus 0.4em\relax
  IEEE, 2003, pp. 2010--2015.

\bibitem{anderson2}
M.~Deghat, B.~Anderson, and Z.~Lin, ``Combined flocking and distance-based
  shape control of multi-agent formations.'' \emph{IEEE Transactions on
  Automatic Control, 61}, pp. 1824--1837, 2016.

\bibitem{sun2015rigid}
Z.~Sun and B.~D. Anderson, ``Rigid formation control systems modelled by double
  integrators: System dynamics and convergence analysis,'' in \emph{2015 5th
  Australian Control Conference (AUCC)}.\hskip 1em plus 0.5em minus 0.4em\relax
  IEEE, 2015, pp. 241--246.

\bibitem{dika}
D.~V. {Dimarogonas} and K.~H. {Johansson}, ``On the stability of distance-based
  formation control,'' in \emph{2008 47th IEEE Conference on Decision and
  Control}, Dec 2008, pp. 1200--1205.

\bibitem{krick2009stabilisation}
L.~Krick, M.~E. Broucke, and B.~A. Francis, ``Stabilisation of infinitesimally
  rigid formations of multi-robot networks,'' \emph{International Journal of
  control}, vol.~82, no.~3, pp. 423--439, 2009.

\bibitem{oh2015survey}
K.-K. Oh, M.-C. Park, and H.-S. Ahn, ``A survey of multi-agent formation
  control,'' \emph{Automatica}, vol.~53, pp. 424--440, 2015.

\bibitem{olfati2006flocking}
R.~Olfati-Saber, ``Flocking for multi-agent dynamic systems: Algorithms and
  theory,'' \emph{IEEE Transactions on automatic control}, vol.~51, no.~3, pp.
  401--420, 2006.

\bibitem{liu2012iterative}
Y.~Liu and Y.~Jia, ``An iterative learning approach to formation control of
  multi-agent systems,'' \emph{Systems \& Control Letters}, vol.~61, no.~1, pp.
  148--154, 2012.

\bibitem{yuan2017formation}
C.~Yuan, S.~Licht, and H.~He, ``Formation learning control of multiple
  autonomous underwater vehicles with heterogeneous nonlinear uncertain
  dynamics,'' \emph{IEEE Transactions on Cybernetics}, vol.~48, no.~10, pp.
  2920--2934, 2017.

\bibitem{yan2021flocking}
T.~Yan, X.~Xu, Z.~Li, and E.~Li, ``Flocking of multi-agent systems with unknown
  nonlinear dynamics and heterogeneous virtual leader,'' \emph{International
  Journal of Control, Automation and Systems}, vol.~19, no.~9, pp. 2931--2939,
  2021.

\bibitem{8870252}
G.~{Wen}, C.~L.~P. {Chen}, and B.~{Li}, ``Optimized formation control using
  simplified reinforcement learning for a class of multiagent systems with
  unknown dynamics,'' \emph{IEEE Transactions on Industrial Electronics},
  vol.~67, no.~9, pp. 7879--7888, 2020.

\bibitem{schilling2019learning}
F.~Schilling, J.~Lecoeur, F.~Schiano, and D.~Floreano, ``Learning vision-based
  flight in drone swarms by imitation,'' \emph{IEEE Robotics and Automation
  Letters}, vol.~4, no.~4, pp. 4523--4530, 2019.

\bibitem{tolstaya2020learning}
E.~Tolstaya, F.~Gama, J.~Paulos, G.~Pappas, V.~Kumar, and A.~Ribeiro,
  ``Learning decentralized controllers for robot swarms with graph neural
  networks,'' in \emph{Conference on robot learning}.\hskip 1em plus 0.5em
  minus 0.4em\relax PMLR, 2020, pp. 671--682.

\bibitem{gama2021graph}
F.~Gama, E.~Tolstaya, and A.~Ribeiro, ``Graph neural networks for decentralized
  controllers,'' in \emph{ICASSP 2021-2021 IEEE International Conference on
  Acoustics, Speech and Signal Processing (ICASSP)}.\hskip 1em plus 0.5em minus
  0.4em\relax IEEE, 2021, pp. 5260--5264.

\bibitem{yang2021distributed}
Z.~Yang, S.~Sosnowski, Q.~Liu, J.~Jiao, A.~Lederer, and S.~Hirche,
  ``Distributed learning consensus control for unknown nonlinear multi-agent
  systems based on gaussian processes,'' \emph{arXiv preprint
  arXiv:2103.15929}, 2021.

\bibitem{Beckers2021}
T.~Beckers, S.~Hirche, and L.~Colombo, ``Safe online learning-based formation
  control of multi-agent systems with gaussian processes,'' in \emph{Proc. of
  the Conference on Decision and Control (to appear)}, 2021.

\bibitem{Ji2021}
T.~Z. Jiahao, L.~Pan, and M.~A. Hsieh, ``Learning to swarm with knowledge-based
  neural ordinary differential equations,'' \emph{arXiv preprint
  arXiv:2109.04927}, 2021.

\bibitem{be2021}
M.~B. Bezcioglu, B.~Lennox, and F.~Arvin, ``Self-organised swarm flocking with
  deep reinforcement learning,'' in \emph{2021 7th International Conference on
  Automation, Robotics and Applications (ICARA)}.\hskip 1em plus 0.5em minus
  0.4em\relax IEEE, 2021, pp. 226--230.

\bibitem{hou2013model}
Z.-S. Hou and Z.~Wang, ``From model-based control to data-driven control:
  Survey, classification and perspective,'' \emph{Information Sciences}, vol.
  235, pp. 3--35, 2013.

\bibitem{rasmussen2006gaussian}
C.~E. Rasmussen and C.~K. Williams, \emph{{Gaussian} processes for machine
  learning}.\hskip 1em plus 0.5em minus 0.4em\relax MIT press Cambridge, 2006,
  vol.~1.

\bibitem{hewing2019cautious}
L.~Hewing, J.~Kabzan, and M.~N. Zeilinger, ``Cautious model predictive control
  using gaussian process regression,'' \emph{IEEE Transactions on Control
  Systems Technology}, vol.~28, no.~6, pp. 2736--2743, 2019.

\bibitem{umlauft:TAC2020}
J.~Umlauft and S.~Hirche, ``Feedback linearization based on gaussian processes
  with event-triggered online learning,'' \emph{IEEE Transactions on Automatic
  Control}, 2020.

\bibitem{beckers2019stable}
T.~Beckers, D.~Kuli{\'c}, and S.~Hirche, ``Stable {G}aussian process based
  tracking control of {E}uler--{L}agrange systems,'' \emph{Automatica}, vol.
  103, pp. 390--397, 2019.

\bibitem{godsil2001algebraic}
C.~Godsil and G.~F. Royle, \emph{Algebraic graph theory}.\hskip 1em plus 0.5em
  minus 0.4em\relax Springer Science \& Business Media, 2001, vol. 207.

\bibitem{asimov}
B.~R. L.~Asimow, ``The rigidity of graphs, ii,'' \emph{Journal of Mathematical
  Analysis and Applications}, vol.~68, no.~1, pp. 171--190, 1979.

\bibitem{aastrom1971system}
K.~J. {\AA}str{\"o}m and P.~Eykhoff, ``System identification—a survey,''
  \emph{Automatica}, vol.~7, no.~2, pp. 123--162, 1971.

\bibitem{wahba1990spline}
G.~Wahba, \emph{Spline models for observational data}.\hskip 1em plus 0.5em
  minus 0.4em\relax SIAM, 1990.

\bibitem{beckers2017stable}
T.~Beckers, J.~Umlauft, D.~Kulic, and S.~Hirche, ``Stable {G}aussian process
  based tracking control of lagrangian systems,'' in \emph{2017 IEEE 56th
  Annual Conference on Decision and Control (CDC)}.\hskip 1em plus 0.5em minus
  0.4em\relax IEEE, 2017, pp. 5180--5185.

\bibitem{srinivas2012information}
N.~Srinivas, A.~Krause, S.~M. Kakade, and M.~W. Seeger, ``Information-theoretic
  regret bounds for {{G}aussian} process optimization in the bandit setting,''
  \emph{IEEE Transactions on Information Theory}, vol.~58, no.~5, pp.
  3250--3265, 2012.

\bibitem{ahn14}
K.~Oh and H.~Ahn, ``Distance-based undirected formation of single-integrator
  and double-integrator modeled agents in $n$-dimensional space,''
  \emph{International Journal of Robust and Nonlinear Control}, pp. 1809--1820,
  2014.

\bibitem{dorfler2011critical}
F.~D{\"o}rfler and F.~Bullo, ``On the critical coupling for kuramoto
  oscillators,'' \emph{SIAM Journal on Applied Dynamical Systems}, vol.~10,
  no.~3, pp. 1070--1099, 2011.

\bibitem{beckers2016equilibrium}
T.~Beckers and S.~Hirche, ``Equilibrium distributions and stability analysis of
  {G}aussian process state space models,'' in \emph{2016 IEEE 55th Conference
  on Decision and Control (CDC)}.\hskip 1em plus 0.5em minus 0.4em\relax IEEE,
  2016, pp. 6355--6361.

\end{thebibliography}
